# Optimal Throughput-Diversity-Delay Tradeoff in MIMO ARQ Block-Fading Channels

Allen Chuang, Albert Guillén i Fàbregas, Lars K. Rasmussen, and Iain B. Collings

### Abstract

In this paper, we consider an automatic-repeat-request (ARQ) retransmission protocol signaling over a block-fading multiple-input, multiple-output (MIMO) channel. Unlike previous work, we allow for multiple fading blocks within each transmission (ARQ round), and we constrain the transmitter to fixed rate codes constructed over complex signal constellations. In particular, we examine the general case of average input-power-constrained constellations as well as the practically important case of finite discrete constellations. This scenario is a suitable model for practical wireless communications systems employing orthogonal frequency division multiplexing techniques over a MIMO ARQ channel. Two cases of fading dynamics are considered, namely short-term static fading where channel fading gains change randomly for each ARQ round, and long-term static fading where channel fading gains remain constant over all ARQ rounds pertaining to a given message. As our main result, we prove that for the block-fading MIMO ARQ channel with discrete input signal constellation satisfying a short-term power constraint, the optimal signal-to-noise ratio (SNR) exponent is given by a modified Singleton bound, relating all the system parameters. To demonstrate the practical significance of the theoretical analysis, we present numerical results showing that practical Singleton-bound-achieving maximum distance separable codes achieve the optimal SNR exponent.

A. Chuang is with the School of Electrical and Information Engineering, University of Sydney (e-mail: achuang@ee.usyd.edu.au).

A. Guillén i Fàbregas was with the Institute for Telecommunications Research, University of South Australia. He is now with the Department of Engineering, University of Cambridge, UK, (e-mail: guillen@ieee.org).

L. K. Rasmussen is with the Institute for Telecommunications Research, University of South Australia (e-mail: lars.rasmussen@unisa.edu.au).

I. B. Collings is with the ICT Centre, CSIRO (e-mail: iain.collings@csiro.au).

This work was supported by the Australian Research Council under ARC grants DP0558861 and RN0459498. The material of this paper has been presented in part at the 2006 IEEE Inf. Theory Workshop, Chengdu, China, Oct. 2006.

## I. Introduction

In 1957 multi-carrier transmission was first proposed by Doelz et al. [1] as a way to increase data rate by transmitting multiple bits streams in parallel over multiple carriers. Originally, multi-carrier transmission was implemented using banks of sinusoidal generators. The use of discrete Fourier transforms for modulation and demodulation was first suggested by Weinstein and Ebert in 1971 [2], significantly reducing implementation complexity, and leading to what we now know as *orthogonal frequency division multiplexing* (OFDM). A review of the development of multi-carrier and OFDM systems can be found in [3].

Almost fifty years after the invention of multi-carrier transmission [1, 2], the use of OFDM has been adopted for broadband wireless communications systems as a means to significantly increase transmission rates [4]. Standards such as IEEE 802.11 (WiFi) [5, 6] and IEEE 802.16 (WiMax) [7, 8] have now been extended to include OFDM techniques. Further improvements of data rate and reliability are promised through the use of multiple transmit and receive antennas [9, 10]. Multiple-input, multiple-output (MIMO) antenna systems are now being introduced into the IEEE 802 standards [6, 8], as well as being integral parts of fourth-generation mobile cellular communication systems proposals [11, 12]. In addition, adaptive coding and modulation, combined with automatic-repeat-request (ARQ) retransmission protocols, are becoming integral parts of data transmission services in the Universal Mobile Telecommunications System (UMTS) [13], and in WiMax [8].

Practical wireless communication systems will therefore soon feature MIMO OFDM modulation with overlaying ARQ protocols. It is thus important to obtain a thorough understanding of the fundamental characteristics of such systems. In this paper, we model a practical point-to-point MIMO OFDM ARQ wireless communication system as a system transmitting signals from a complex signal constellation over a block-fading MIMO ARQ channel. In the following subsections, we first review prior art and technical concepts relevant to our work. We then formulate our problem and summarize contributions, before outlining the organization and defining notation of the paper.

### A. Prior Art

*1) Fundamental Tradeoff:* The work of Teletar [14], and Foschini and Gans [15], has inspired a flurry of research activities in MIMO antenna systems for wireless communications. Previously, multiple-antenna systems were primarily used for providing receiver diversity, thus combatting random amplitude fluctuations due to fading [16]. In contrast, the prevailing thesis for MIMO systems is that fading can increase channel capacity by providing a set of well-behaved parallel channels [14, 15]. In fact, in the high signal-to-noise (SNR) regime it has been shown that the capacity of a channel with $N_t$ transmit antennas, $N_r$ receive antennas, and independent, identical distributed (i.i.d.) complex Gaussian channel gains between each antenna pair is given by

$$\mathsf{C}(\mathrm{SNR}) = \min\{N_t, N_r\} \log \mathrm{SNR} + O(1),$$

suggesting that capacity increases linearly with the minimum number of transmit and receive antennas. Therefore, the use of multiple-antenna systems can improve both reliability and data rate, when transmitting over a quasi-static MIMO channel where channel gains are i.i.d. complex Gaussian and fixed during the transmission.

Zheng and Tse described the fundamental tradeoff between diversity gain and multiplexing gain[1] for quasi-static MIMO channels in the high SNR regime in [17], assuming Gaussian distributed input signals. The fundamental tradeoff developed in [17] has since become a benchmark for the performance evaluation of space-time coding schemes, and the corresponding framework has become a preferred approach for characterizing classes of MIMO channels. For example, in [18] the fundamental diversity-multiplexing-delay tradeoff is characterized for the MIMO ARQ channel, and the fundamental diversity-multiplexing tradeoff for MIMO channels with resolution-constrained feedback is determined in [19], both under the assumption that Gaussian distributed input signals are used.

*2) OFDM and the Block-Fading Channel:* The block-fading channel model was introduced in [20], with the purpose of modelling slowly varying fading channels where the duration of a block-fading period is determined by the channel coherence time. Within a block-fading period, the

---

[1]The diversity gain (or SNR exponent) is defined as

$$d \triangleq -\lim_{\mathrm{SNR} \to \infty} \frac{\log P_e(\mathrm{SNR})}{\log \mathrm{SNR}},$$

where $P_e(\mathrm{SNR})$ denotes the probability that the transmitted message is decoded incorrectly. The multiplexing gain is defined as

$$r_m \triangleq \lim_{\mathrm{SNR} \to \infty} \frac{R(\mathrm{SNR})}{\log \mathrm{SNR}},$$

where $R(\mathrm{SNR})$ is the code rate. The multiplexing gain essentially quantifies how close the code rate is to the capacity of a single-input single-output link at high SNR [17].

channel fading gain remains constant, while between periods the channel gains change randomly according to a fading distribution. In this setting, transmission typically extends over multiple block-fading periods. A thorough treatment of fading channels is found in [21].

The block-fading channel model is a reasonable model for OFDM transmission over frequency-selective wireless channels, as an OFDM system is typically designed such that each sub-carrier experiences flat fading. Despite its simplicity, the model captures important aspects of OFDM modulation over frequency-selective fading channels and proves useful for developing coding design criteria.

The definition of multiplexing gain, fundamental in the formulation presented in [17, 18], relies on coding schemes with transmission rates that increase linearly with the logarithm of the SNR. Non-zero multiplexing gains can only be achieved with continuous input constellations or discrete constellations with cardinalities scaling with the SNR. From a practical perspective, it is desirable to operate at a fixed code rate and deal with small alphabet sizes. We are therefore interested in the performance of such practical schemes, which effectively operates at zero multiplexing gain. Under this scenario, the general diversity-multiplexing tradeoff can only provide a coarse characterization of the *rate-diversity* tradeoff. The rate-diversity tradeoff for fixed-rate space-time codes constructed over discrete signal constellations, and transmitted across a quasi-static MIMO channel, was presented in [22].

Union-bound arguments [23] and error exponent calculations [24] were used to show that the diversity gain of a block-fading channel with an arbitrary, but fixed number of fading blocks, fixed code rate, and a discrete input signal constellation, is described by a modified version of the Singleton bound [25]. The same problem is considered in [26], where outage probability arguments are used to formally prove that the optimal rate-diversity tradeoff is indeed the modified form of the Singleton bound presented in [24, 25], which is achieved using maximum distance separable (MDS) codes.

The block-fading ARQ channel model has recently been considered in [27, 28] for discrete input signal constellations. In [27] the Singleton bound is presented as an upper bound to the SNR exponent, while the optimality of the Singleton bound is formally proven for the ARQ case in [28]. In [28] it is also demonstrated that asymptotically optimal throughput can be achieved by MDS codes.

## B. Problem Formulation and Contributions

In this paper, we consider an ARQ system signaling over a block-fading MIMO channel with $L$ maximum number of allowable ARQ rounds and $B$ fading blocks per ARQ round. In contrast to the work in [17, 18], we allow for multiple fading blocks within each transmission (ARQ round), and we constrain the transmitter to fixed rate codes constructed over complex signal constellations. In particular, we examine the general case of average input-power-constrained constellations as well as the practically important case of discrete constellations of finite cardinality. The receiver is able to generate a finite number of one-bit repeat-requests, subject to a latency constraint, whenever an error is detected in the decoded message. A maximum of $L$ transmissions pertaining to each information message is allowed.

As in [18], we consider two cases of fading statistics; for the short-term static fading case, the channel fading gains change randomly for each ARQ round, while for the long-term static fading case, the channel fading gains remain constant over all ARQ rounds pertaining to a given message, but change randomly for each message and corresponding suite of ARQ rounds. This scenario is a suitable model for practical wireless communications systems employing OFDM modulation over a MIMO ARQ channel.

The main focus of our work is to derive the optimal tradeoff between throughput, diversity gain and delay of ARQ schemes signaling over block-fading MIMO channels. In particular, we show that the tradeoff highlights the roles of the complex-plane signal constellation, the rate of the first ARQ round $R_1$, the maximum number of ARQ rounds $L$, and the number of fading blocks per ARQ round $B$.

As a first result, we prove that for the block-fading MIMO ARQ channel with the input constellation satisfying a short-term power constraint, the optimal SNR exponent is given by $N_t N_r L B$ for short-term static fading and $N_t N_r B$ for long-term static fading, which is achieved by Gaussian codes of any positive rate. This is, however, not the case with discrete signal constellations. In order to attain full diversity the signal constellations must feature certain properties. In general, due to the discrete nature of these signal sets, a tradeoff between rate, diversity and delay arises.

As our main result, we prove that for the block-fading MIMO ARQ channel with discrete input signal constellation of cardinality $2^{QN_t}$ satisfying a short-term power constraint, the optimal

SNR exponent is given by a modified Singleton bound, relating all the system parameters. Note, however, that modulating across all fading blocks increases the dimensionality of the decoding problem by a factor of $B$ [29]. For further flexibility in terms of decoding complexity, we consider the case where modulation is performed over a number $1 \leq M \leq B$ of fading blocks, such that $B = MD$. The resulting optimal SNR exponent is then expressed as

$$d_D^\star(R_1) = \begin{cases} MN_tN_r \left( 1 + \left\lfloor \dfrac{LB}{M} \left( 1 - \dfrac{R_1}{LQN_t} \right) \right\rfloor \right) & \text{for short-term static fading} \\[3mm] MN_tN_r \left( 1 + \left\lfloor \dfrac{B}{M} \left( 1 - \dfrac{R_1}{LQN_t} \right) \right\rfloor \right) & \text{for long-term static fading} \end{cases} \quad (1)$$

The expression in (1) implies that as the target rate $R_1$ increases, the achievable optimal diversity order $d_D^\star(R_1)$ decreases in steps. Our main result generalizes the result of [22] for the quasi-static MIMO channel to the ARQ block-fading case with encoding across $M$ fading blocks.

Directly following from the results, we demonstrate that while the optimal SNR exponent of the system is an increasing function of the maximum number of allowed ARQ rounds $L$, the throughput of the system becomes independent of $L$ for sufficiently high SNR, and is determined by the rate of the first ARQ round. We therefore denote our main result as the *optimal throughput-diversity-delay tradeoff*. This result provides strong incentive to use ARQ as a way to increase reliability without suffering code rate penalties.

To demonstrate the practical coding aspects of our results, some examples are presented with corresponding error rate and throughput performances. The diversity tradeoff function can be viewed as a modified version of the Singleton bound [25], which naturally leads us to investigate the role of Singleton-bound-achieving MDS codes. Our examples illustrate that the optimal SNR exponent can be achieved with practical MDS coding schemes.

*C. Organization and Notation*

The paper is organized as follows. In Section II we define the system model, and in Section III we review relevant ARQ performance measures, namely, error probability, throughput and average latency. In Section IV we review the concepts of information accumulation and outage probability, while the main theorems of the paper, detailing the throughput-diversity-delay tradeoff, are presented in Section V. A thorough discussion is included in Section V, where the results are interpreted and related to existing results in the literature. To demonstrate the practical relevance of the results, numerical examples are included in Section VI, showing that MDS codes achieve the tradeoff. Concluding remarks are summarized in Section VII, while the details of the proofs have been collected in the appendices.

The following notation is used in the paper. Sets are denoted by calligraphic fonts with the complement denoted by superscript $c$. The exponential equality $f(z) \doteq z^d$ indicates that $\lim_{z \to \infty} \frac{\log f(z)}{\log z} = d$. The exponential inequality $\dot{\leq}, \dot{\geq}$ are similarly defined. $\succ$ and $\prec$ denote component-wise inequality of $>$ and $<$, respectively. $\mathbf{I}$ denotes the identity matrix, vector/matrix transpose is denoted by $'$ (e.g. $\mathbf{v}'$) and $\| \cdot \|_F$ is the Frobenius norm. $\mathbb{1}\{\cdot\}$ is the indicator function, and $\lceil x \rceil$ ($\lfloor x \rfloor$) denotes the smallest (largest) integer greater (smaller) than $x$.

## II. System Model

In this section we describe the block-fading MIMO ARQ channel model and coded modulation schemes under consideration.

### A. Channel Model

Consider a block-fading MIMO ARQ system with $N_t$ transmit antennas and $N_r$ receive antennas. We investigate the use of a simple stop-and-wait ARQ protocol where the maximum number of ARQ rounds is denoted by $L$. Each ARQ round consists of $B$ independent block-fading periods, each of length $T$ (coherence time/bandwidth) in channel uses. Hence each ARQ round spans $BT$ channel uses. Figure 1 shows the overall system model. We write the received signal at the $b$th block and $\ell$th ARQ round as

$$\mathbf{Y}_{\ell,b} = \sqrt{\frac{\rho}{N_t}} \mathbf{H}_{\ell,b} \mathbf{X}_{\ell,b} + \mathbf{W}_{\ell,b}, \tag{2}$$

where $\mathbf{X}_{\ell,b} \in \mathbb{C}^{N_t \times T}, \mathbf{Y}_{\ell,b}, \mathbf{W}_{\ell,b} \in \mathbb{C}^{N_r \times T}$ and $\mathbf{H}_{\ell,b} \in \mathbb{C}^{N_r \times N_t}$ denote the transmitted signal matrix, received signal matrix, the noise matrix and the channel fading gain matrix, respectively. We define $\mathbf{x}_{\ell,b,t} \in \mathbb{C}^{N_t}$ as the vectors containing the transmitted symbols of each antenna at ARQ round $\ell$, block $b$ and time $t$, which are such that $\mathbf{X}_{\ell,b} = [\mathbf{x}_{\ell,b,1}, \ldots, \mathbf{x}_{\ell,b,T}]$.

Both the elements of the channel fading gain matrix $\mathbf{H}_{\ell,b}$ and the elements of the noise matrix $\mathbf{W}_{\ell,b}$ are assumed i.i.d. zero mean complex circularly symmetric complex Gaussian with variance $\sigma^2 = 0.5$ per dimension. We assume perfect receiver-side channel state information (CSI), namely, the channel coefficients are assumed to be perfectly known to the receiver.

We obtain the *long-term static* model of [18] by letting $\mathbf{H}_{\ell,b} = \mathbf{H}_{\ell',b}$ for all $\ell \neq \ell'$ in (2), namely, all ARQ rounds undergo the same MIMO block-fading channel. This models well a slowly varying MIMO OFDM ARQ system with $B$ subcarriers or $B$ groups of correlated subcarriers. On the other hand, when the matrices $\mathbf{H}_{\ell,b}$ are i.i.d. from block to block and from ARQ round to ARQ round, (2) corresponds to the *short-term static* model of [18]. In order to keep the presentation general, and since (2) encompasses both models, we will index the channel matrices according to ARQ round and block as in the short-term static model. We will outline the changes for the long-term static model whenever necessary.

Therefore, the channel model *corresponding to ARQ round $\ell$* becomes

$$\mathbf{Y}_\ell = \sqrt{\frac{\rho}{N_t}} \mathbf{H}_\ell \mathbf{X}_\ell + \mathbf{W}_\ell, \tag{3}$$

where

$$\mathbf{Y}_\ell = \left[ \mathbf{Y}'_{\ell,1}, \ldots, \mathbf{Y}'_{\ell,B} \right]' \in \mathbb{C}^{BN_r \times T}$$

$$\mathbf{X}_\ell = \left[ \mathbf{X}'_{\ell,1}, \ldots, \mathbf{X}'_{\ell,B} \right]' \in \mathbb{C}^{BN_t \times T}$$

$$\mathbf{W}_\ell = \left[ \mathbf{W}'_{\ell,1}, \ldots, \mathbf{W}'_{\ell,B} \right]' \in \mathbb{C}^{BN_t \times T}$$

$$\mathbf{H}_\ell = \mathrm{diag}\left( \mathbf{H}_{\ell,1}, \ldots, \mathbf{H}_{\ell,B} \right) \in \mathbb{C}^{BN_r \times BN_t}.$$

One channel use of the equivalent model (3) corresponds to $BT$ channel uses of the *real* channel (2). In a similar way to the previous model, we define the vectors $\mathbf{x}_{\ell,t} \in \mathbb{C}^{BN_t}$ for $t = 1, \ldots, T$ as

$$\mathbf{X}_\ell = \left[ \mathbf{x}_{\ell,1}, \ldots, \mathbf{x}_{\ell,T} \right] \in \mathbb{C}^{BN_t \times T}.$$

The receiver attempts to decode following the reception of an ARQ round. If the received codeword can be decoded, the receiver sends back a one-bit acknowledgement signal to the transmitter via a zero-delay and error-free feedback link. The transmission of the current codeword ends immediately following the acknowledgment signal and the transmission of the next message in the queue starts. If an error is detected in the received codeword before the $L$th ARQ round, then the receiver requests another ARQ round by sending back a one-bit negative acknowledgment along the perfect feedback path. However, a decision must be made at the end of the $L$th ARQ round regardless of whether errors are detected.

In general, the optimal ARQ decoder makes use of all available coded blocks and corresponding channel state information up to the current ARQ round in the decoding process. This leads to the concept of information accumulation, where individual ARQ rounds are combined, along with any other side information. We hence introduce the ARQ channel model *up to the $\ell$th ARQ round*, completely analogous to (2), but allowing for a more concise notation. In particular, we have that

$$\widetilde{\mathbf{Y}}_\ell = \sqrt{\frac{\rho}{N_t}} \widetilde{\mathbf{H}}_\ell \widetilde{\mathbf{X}}_\ell + \widetilde{\mathbf{W}}_\ell, \tag{4}$$

where

$$\widetilde{\mathbf{Y}}_\ell = [\mathbf{Y}_1', \ldots, \mathbf{Y}_\ell']' \in \mathbb{C}^{\ell B N_r \times T},$$

$$\widetilde{\mathbf{X}}_\ell = [\mathbf{X}_1', \ldots, \mathbf{X}_\ell']' \in \mathbb{C}^{\ell B N_t \times T},$$

$$\widetilde{\mathbf{W}}_\ell = [\mathbf{W}_1', \ldots, \mathbf{W}_\ell']' \in \mathbb{C}^{\ell B N_r \times T},$$

$$\widetilde{\mathbf{H}}_\ell = \mathrm{diag}\,(\mathbf{H}_1, \ldots, \mathbf{H}_\ell) \in \mathbb{C}^{\ell B N_r \times \ell B N_t}.$$

That is, $\widetilde{\mathbf{Y}}_\ell$, $\widetilde{\mathbf{X}}_\ell$ and $\widetilde{\mathbf{W}}_\ell$ are simply collections of the received, code and noise matrices, respectively, available at the end of the $\ell$th ARQ round, concatenated into block column matrices. The new channel matrix $\widetilde{\mathbf{H}}_\ell \in \mathbb{C}^{\ell B N_r \times \ell B N_t}$ is a block diagonal matrix with the diagonal blocks composed of the respective channel state during each block-fading period up to ARQ round $\ell$. In the case of long-term static model, $\widetilde{\mathbf{H}}_\ell = \mathrm{diag}\,(\underbrace{\mathbf{H}, \ldots, \mathbf{H}}_{\ell \text{ times}})$. Note that a channel use of the equivalent model (4) corresponds to $\ell B T$ channel uses of the *real* channel (2).

## B. Encoding

In this section we discuss the specific construction of the space-time ARQ codewords. The information message $m$ to be transmitted is passed through a space-time coded modulation encoder with codebook $\mathcal{C} \subset \mathbb{C}^{L B N_t \times T}$ and code rate $R_0$, where $R_0 \triangleq \frac{R_1}{L}$ and

$$R_1 \triangleq \frac{1}{BT} \log_2 |\mathcal{C}|$$

is the code rate of the first ARQ round. Therefore, $|\mathcal{C}| = 2^{R_0 L B T}$ and $m \in \mathcal{M}$, where $\mathcal{M} \triangleq \{1, 2, \ldots, 2^{R_0 L B T}\}$ is the set of possible information messages. We denote the codeword corresponding to information message $m$ by $\mathbf{X}(m)$. The rate $R_0$ codeword can be partitioned into a sequence of $LB$ space-time coded matrices, denoted $\mathbf{X}_{\ell,b} \in \mathbb{C}^{N_t \times T}$. According to the previously described model, we have that

$$\mathbf{X}(m) = \left[\mathbf{X}'_1(m), \ldots, \mathbf{X}'_L(m)\right]'$$
$$= \left[\mathbf{X}'_{1,1}(m), \ldots, \mathbf{X}'_{1,B}(m), \ldots, \mathbf{X}'_{L,1}(m), \ldots, \mathbf{X}'_{L,B}(m)\right]' \in \mathbb{C}^{LBN_t \times T}$$

We consider a *short term* average power constraint, namely, the transmitted codewords are normalized in energy such that $\forall \mathbf{X} \in \mathcal{C}, \frac{1}{LBT}\mathbb{E}[\|\mathbf{X}\|_F^2] = N_t$. Therefore, together with the model assumptions in the previous section, $\rho$ in (2), (3) and (4) represents the average SNR per receive antenna.

In this paper we analyze space-time coded modulation schemes constructed over discrete signal sets. In particular, we consider that $\mathcal{C}$ is obtained as the concatenation of a *classical coded modulation scheme* $\mathcal{C}_\mathcal{Q} \subseteq \mathcal{Q}^{LBTN_t}$ constructed over a complex-plane signal set $\mathcal{Q} = \{q_1, \ldots, q_{|\mathcal{Q}|}\} \subset \mathbb{C}$ [30] with a unit rate linear dispersion space-time modulator [31]. Let $\mathbf{c}_\mathcal{Q} \in \mathcal{C}_\mathcal{Q}$ denote a codeword of $\mathcal{C}_\mathcal{Q}$ of length $LBTN_t$ and $Q = \log_2 |\mathcal{Q}|$ the number of bits conveyed in one symbol of $\mathcal{Q}$, namely, $|\mathcal{Q}| = 2^Q$. Since the linear dispersion space-time modulator has unit rate we have that $0 \leq R_1 \leq N_t Q L$.

To allow for a general case, we consider that the linear dispersion space-time modulator spreads the symbols of $\mathbf{c}_\mathcal{Q}$ over the $N_t$ transmit antennas and the $B$ fading blocks. In particular, we consider that the codewords $\mathbf{c}_\mathcal{Q}$ of $\mathcal{C}_\mathcal{Q}$, of length $LBTN_t$ are partitioned into $L$ vectors of length $BTN_t$ each, denoted by $\mathbf{c}_{\mathcal{Q},\ell} \in \mathcal{Q}^{BTN_t}$ such that $\mathbf{c}_\mathcal{Q} = \left[\mathbf{c}'_{\mathcal{Q},1}, \ldots, \mathbf{c}'_{\mathcal{Q},L}\right]'$. For every $\ell = 1, \ldots, L$, the vectors $\mathbf{c}_{\mathcal{Q},\ell}$ are multiplied by the unit rate generator matrix of the linear dispersion space-time modulator $\mathbf{R} \in \mathbb{C}^{BTN_t \times BTN_t}$ to form

$$\mathbf{x}_\ell = \mathbf{R}\,\mathbf{c}_{\mathcal{Q},\ell} \tag{5}$$

where $\mathbf{x}_\ell = \text{vec}(\mathbf{X}_\ell) \in \mathbb{C}^{BN_tT}$ is the vector representation of the portion of codeword of $\mathcal{C} \in \mathcal{C}_\mathcal{Q}$ transmitted at ARQ round $\ell$. Without any loss in generality we consider that $\mathbf{R}$ is a rotation matrix [32–35], i.e., $\mathbf{R}$ is unitary [36]. Note that introduction of the linear dispersion space-time modulator rotation matrix $\mathbf{R}$ increases the decoding complexity compared to the unrotated case where $\mathbf{R} = \mathbf{I}$. This is due to the fact that now the components of $\mathbf{x}_\ell$ depend on each other, since $\mathbf{R}$ induces a change of the reference axis for detection [32–35]. This implies that the detection problem is of dimension $BTN_t$.

To allow for further flexibility, we consider the case where the linear dispersion space-time modulator spreads the symbols of $\mathbf{c}_\mathcal{Q} \in \mathcal{C}_\mathcal{Q}$ over the $N_t$ transmit antennas and a number $1 \leq M \leq B$ of fading blocks, such that

$$D \triangleq \frac{B}{M}$$

is an integer representing the number of rotations used in an ARQ round. In this case, we have that the rotation matrix $\mathbf{R}$ becomes block-diagonal, namely

$$\mathbf{R} = \mathrm{diag}\big(\underbrace{\mathbf{R}_M, \ldots, \mathbf{R}_M}_{D \text{ times}}\big) \tag{6}$$

where $\mathbf{R}_M \in \mathbb{C}^{MN_tT \times MN_tT}$ is the rotation matrix of dimension $MN_tT \times MN_tT$. According to (6) we can define $\widehat{\mathbf{x}}_{\ell,d} \in \mathbb{C}^{MN_tT}$, such that $\widehat{\mathbf{x}}_\ell = [\widehat{\mathbf{x}}_{\ell,1}, \ldots, \widehat{\mathbf{x}}_{\ell,D}]'$. We define the multidimensional constellation $\mathcal{X}_M$ as

$$\mathcal{X}_M \triangleq \big\{ \mathbf{x} \in \mathbb{C}^{MN_tT} \; : \; \forall \mathbf{c} \in \mathcal{Q}^{MN_tT}, \; \mathbf{x} = \mathbf{R}_M\,\mathbf{c} \big\} \tag{7}$$

Due to the block-diagonal structure of $\mathbf{R}$, the detection problem reduces to $D$ detection problems over $\mathcal{X}_M$ each of dimension $MTN_t$. This formulation encompasses many cases of interest, as for example the unrotated case, for which $\mathbf{R} = \mathbf{I}$, the general threaded algebraic space-time (TAST) modulation structure for MIMO block-fading channels [29], or perfect space-time modulation [37]. As we shall see in Section V, the parameter $M$ plays a key role in the reliability of the overall system. Intuitively, the larger $M$, the larger the space-time symbol spreading, and hence, the larger the diversity [29]. On the other hand, using large $M$ implies larger decoding complexity, as the detection problem is exponential in $M$. Using the previous discussion, we introduce the following equivalent channel matrix

$$\widehat{\mathbf{H}}_{\ell,d} = \mathrm{diag}\big(\underbrace{\mathbf{H}_{\ell,(d-1)M+1}, \ldots, \mathbf{H}_{\ell,(d-1)M+1}}_{T \text{ times}}, \ldots, \underbrace{\mathbf{H}_{\ell,dM}, \ldots, \mathbf{H}_{\ell,dM}}_{T \text{ times}}\big) \in \mathbb{C}^{MN_rT \times MN_tT} \tag{8}$$

for $d = 1, \ldots, D$. These matrices correspond to the channels seen by rotation $d$ within ARQ round $\ell$. The equivalent channel defined by (8) induces the following channel model

$$\widehat{\mathbf{y}}_{\ell,d} = \sqrt{\frac{\rho}{N_t}}\widehat{\mathbf{H}}_{\ell,d}\widehat{\mathbf{x}}_{\ell,d} + \widehat{\mathbf{w}}_{\ell,d} \tag{9}$$

where $\widehat{\mathbf{x}}_{\ell,d} \in \mathbb{C}^{MN_tT}, \widehat{\mathbf{y}}_{\ell,d}, \widehat{\mathbf{w}}_{\ell,d} \in \mathbb{C}^{MN_rT}$ are the corresponding input, output and noise vectors. This model describes the relationship between the output of one of the $D$ rotations of the linear dispersion space-time modulator and the output of the channel. One use of channel (9) corresponds to $MT$ uses of the *real* model (2).

## C. Decoding

We will make use of the ARQ decoder proposed in [18], which behaves as a typical set decoder for the first $L-1$ ARQ round and finally performs ML decoding at the last ARQ round. The decoding function at ARQ round $\ell$, for $\ell = 1, \ldots, L-1$, denoted $\psi_\ell(\widetilde{\mathbf{Y}}_\ell, \widetilde{\mathbf{H}}_\ell)$, gives the following output

$$\psi_\ell(\widetilde{\mathbf{Y}}_\ell, \widetilde{\mathbf{H}}_\ell) = \begin{cases} \hat{m} & \text{if } \widetilde{\mathbf{X}}(\hat{m}) \text{ is the unique codeword in } \mathcal{C} \text{ jointly typical with } \widetilde{\mathbf{Y}}_\ell \text{ given } \widetilde{\mathbf{H}}_\ell \\ 0 & \text{otherwise,} \end{cases}$$

$$(10)$$

which implies that message index $\psi_\ell(\widetilde{\mathbf{Y}}_\ell, \widetilde{\mathbf{H}}_\ell) = \hat{m} \in \mathcal{M}$ whenever the received matrix can be decoded and $\psi_\ell(\widetilde{\mathbf{Y}}_\ell, \widetilde{\mathbf{H}}_\ell) = 0$ whenever errors are detected.

## III. ARQ PERFORMANCE METRICS

In this section we introduce a few performance metrics relevant to ARQ systems, namely, the error probability, average latency and throughput. For ease of notation, we define three relevant decoder events as follows. Let,

$$\mathcal{D}_\ell \triangleq \left\{ \psi_1(\widetilde{\mathbf{Y}}_1, \widetilde{\mathbf{H}}_1) = 0, \ldots, \psi_\ell(\widetilde{\mathbf{Y}}_\ell, \widetilde{\mathbf{H}}_\ell) = 0 \right\}$$

denote the event of error detection up to and including ARQ round $\ell$, let

$$\mathcal{A}_\ell \triangleq \left\{ \bigcup_{\hat{m} \neq 0} \psi_\ell(\widetilde{\mathbf{Y}}_\ell, \widetilde{\mathbf{H}}_\ell) = \hat{m} \right\},$$

denote the event of decoding a valid message at ARQ round $\ell$, and let

$$\mathcal{E}_\ell \triangleq \left\{ \bigcup_{\hat{m} \neq m} \psi_\ell(\widetilde{\mathbf{Y}}_\ell, \widetilde{\mathbf{H}}_\ell) = \hat{m} \right\}$$

denote the event of a decoding error at ARQ round $\ell$, given that message $m$ was transmitted.

Based on the events defined above, the probability of error $P_e(\rho)$ is given by

$$P_e(\rho) = \mathbb{E}\left[ \underbrace{\Pr\left(\mathcal{A}_1, \mathcal{E}_1\right) + \sum_{\ell=2}^{L-1} \Pr\left(\mathcal{D}_{\ell-1}, \mathcal{A}_\ell, \mathcal{E}_\ell\right)}_{\text{undetected errors}} + \underbrace{\Pr\left(\mathcal{D}_{L-1}, \mathcal{E}_L\right)}_{\text{ML decoding errors}} \right],$$

$$(11)$$

where the expectation is with respect to the joint distribution of the fading gain matrix and received signal matrix. From the error expression in (11) it is clear that the ARQ decoder suffers from *undetected errors* and *ML decoding errors*. Undetected errors occur during ARQ rounds $\ell = 1 \ldots L-1$ and reflects the inability of the decoder to identify erroneous frames. ML decoding errors occur at the last ARQ round and reflects the inability of the decoder to resolve atypical channel and noise realizations. We shall see later that the probability of undetected errors can be made arbitrarily small using appropriate codebooks, leaving ML decoding errors to dominate the error probability. In terms of error probability, the effectiveness of an ideal ARQ decoder is therefore almost exclusively limited by the error probability at the last ARQ round.

The expected latency $\kappa$ of the system is determined by the probability of error detection, and it is given by

$$\kappa = 1 + \sum_{\ell=1}^{L-1} \Pr\left(\mathcal{D}_\ell\right), \tag{12}$$

where $\kappa$ is expressed in terms of number of ARQ rounds. The corresponding transmit throughput of the system in terms of the average effective code rate is simply obtained by

$$\eta(R_1, L) = \frac{R_1}{1 + \sum_{\ell=1}^{L-1} \Pr(\mathcal{D}_\ell)}, \tag{13}$$

where $\eta(R_1, L)$ is expressed in bits per channel use[2].

## IV. Information Accumulation and Outage Probability

In this section, we expand on the idea of mutual information accumulation in ARQ systems as well as introduce the commonly used concept of information outage.

The instantaneous input-output mutual information of the channel (4) up to ARQ round $\ell$, for the channel realization $\widetilde{\mathbf{H}}_\ell = \widetilde{\mathbf{G}}_\ell$ can be written as

$$I\left(\rho | \widetilde{\mathbf{G}}_\ell\right) \triangleq \frac{1}{T} I(\widetilde{\mathbf{X}}_\ell \, ; \, \widetilde{\mathbf{Y}}_\ell \mid \widetilde{\mathbf{H}}_\ell = \widetilde{\mathbf{G}}_\ell). \tag{14}$$

$$= \frac{1}{T} \sum_{k=1}^{\ell} I(\rho | \mathbf{G}_\ell) \tag{15}$$

where $I(\rho | \mathbf{G}_\ell)$ is the instantaneous input-output mutual information corresponding to ARQ round $\ell$. Following (15) we will refer to $I\left(\rho | \widetilde{\mathbf{G}}_\ell\right)$ as the *accumulated* mutual information up to ARQ round $\ell$. The accumulated mutual information $I\left(\rho | \widetilde{\mathbf{G}}_\ell\right)$ measures the normalized mutual information between the accumulated received matrix $\widetilde{\mathbf{Y}}_\ell$ and the coded blocks $\widetilde{\mathbf{X}}_\ell$, given the instantaneous channel state matrix $\widetilde{\mathbf{G}}_\ell$. Since $\widetilde{\mathbf{G}}_\ell$ is a random matrix, $I(\rho|\widetilde{\mathbf{G}}_\ell)$ is a non-negative random variable. Further, from (15) it is clear that the accumulated mutual information is an increasing function of the ARQ round index $\ell$, for a given realization of $\widetilde{\mathbf{G}}_\ell$.

---

[2]Note that our definition of transmit throughput here is purely a measure of the average code rate at the sender's side, as it does not take into account whether or not messages are correctly decoded at the receiver's side.

Following [38, Lemma 1], we get that for $|\mathcal{M}| = 2^{R_1 BT}$, there exists a codebook $\mathcal{C}$ such that the conditional probability of error $P_e(\rho|\widetilde{\mathbf{G}}_\ell) < \epsilon$ for any $\epsilon > 0$ whenever the accumulated instantaneous mutual information satisfies $I(\rho|\widetilde{\mathbf{G}}_\ell) \geq R_1$ for any $\ell = 1, \ldots, L$, provided that the block length $\ell BT$ is sufficiently large. We hence define information outage as the event that occurs when the accumulated mutual information is below $R_1$, namely

$$\mathcal{O}_\ell \triangleq \left\{ \widetilde{\mathbf{G}}_\ell \in \mathbb{C}^{\ell BT N_r \times \ell BT N_t} : I(\rho|\widetilde{\mathbf{G}}_\ell) < R_1 \right\}. \tag{16}$$

For any finite $B$ and $L$, the channel defined in (4) is not information stable and the channel capacity in the strict Shannon sense is zero [39], since the probability of the outage event is nonzero. The corresponding outage probability is defined as [20, 21]

$$P_{\text{out}}(\rho, \ell, R_1) \triangleq \Pr\left(\mathcal{O}_\ell\right) \tag{17}$$

$$= \Pr\left(I(\rho|\widetilde{\mathbf{G}}_\ell) < R_1\right). \tag{18}$$

The accumulated mutual information $I(\rho|\widetilde{\mathbf{G}}_\ell)$, and hence the corresponding outage probability, depends on the SNR $\rho$ and the input distribution $P_{\mathbf{X}}(\mathbf{X})$ with the constraint that $\frac{1}{LBT}\mathbb{E}[\|\mathbf{X}\|_F^2] = N_t$. When no other constraints are imposed on the input distribution, the input distribution that maximizes $I(\rho|\widetilde{\mathbf{G}}_\ell)$ and therefore minimizes $P_{\text{out}}(\rho, \ell, R_1)$ is the Gaussian distribution, namely, the entries of $\mathbf{X}$ are i.i.d. complex circularly symmetric random variables with zero mean and unit variance. This leads to,

$$I(\rho|\mathbf{G}_\ell) = \frac{1}{B} \log_2 \det\left(\mathbf{I} + \frac{\rho}{N_t}\mathbf{G}_\ell\mathbf{G}_\ell^\dagger\right) \tag{19}$$

$$= \frac{1}{B} \sum_{b=1}^{B} \log_2 \det\left(\mathbf{I} + \frac{\rho}{N_t}\mathbf{G}_{\ell,b}\mathbf{G}_{\ell,b}^\dagger\right). \tag{20}$$

In practice, Gaussian codebooks are not feasible, and we will resort to discrete signal constellations. In this work, we are mostly interested in studying the role of the discrete nature of practical constellations, and the impact this further system constraint has on the outage probability. In particular, we can write the mutual information for the scheme described in Section II-B as,

$$I(\rho|\mathbf{G}_\ell) = \frac{1}{D}\sum_{d=1}^{D} I(\rho|\widehat{\mathbf{G}}_{\ell,d}) \tag{21}$$

where

$$I(\rho|\widehat{\mathbf{G}}_{\ell,d}) = \frac{\log_2|\mathcal{X}_M|}{MT} - \frac{1}{MT}\mathbb{E}_{\mathbf{x},\mathbf{w}}\left[\log_2\left(\sum_{\mathbf{x}'\in\mathcal{X}_M} e^{-\left\|\sqrt{\frac{\rho}{N_t}}\widehat{\mathbf{G}}_{\ell,d}(\mathbf{x}-\mathbf{x}')+\mathbf{w}\right\|^2+\|\mathbf{w}\|^2}\right)\right] \tag{22}$$

$$= QN_t - \frac{1}{(MT)^2 QN_t}\sum_{\mathbf{x}\in\mathcal{X}_M}\mathbb{E}_{\mathbf{w}}\left[\log_2\left(1+\sum_{\mathbf{x}'\neq\mathbf{x}} e^{-\left\|\sqrt{\frac{\rho}{N_t}}\widehat{\mathbf{G}}_{\ell,d}(\mathbf{x}-\mathbf{x}')+\mathbf{w}\right\|^2+\|\mathbf{w}\|^2}\right)\right] \tag{23}$$

is the input-output mutual information corresponding to the realization $\widehat{\mathbf{H}}_{\ell,d} = \widehat{\mathbf{G}}_{\ell,d}$ given in (8) of the channel described in (9), assuming a uniform distribution over the $MN_tT$ multidimensional constellation $\mathcal{X}_M$ defined in (7). Since $0 \leq I(\rho|\widehat{\mathbf{G}}_{\ell,d}) \leq QN_t$, it is not difficult to show that (21) can be bounded as follows

$$I(\rho|\mathbf{G}_\ell) \leq \frac{1}{D}\sum_{d=1}^{D}\min\left\{QN_t, \frac{1}{M}\sum_{m=1}^{M}\log_2\det\left(\mathbf{I}+\frac{\rho}{N_t}\mathbf{G}_{\ell,(d-1)M+m}\mathbf{G}^\dagger_{\ell,(d-1)M+m}\right)\right\}. \tag{24}$$

This relationship will prove useful in proving our main results.

## V. Throughput-Diversity-Delay Tradeoff

In this section, we derive the optimal tradeoff between throughput, diversity gain and delay of ARQ schemes signaling over MIMO block-fading channels. In particular, we show that the tradeoff highlights the roles of the complex-plane signal constellation through $Q$, the rate of the first ARQ round $R_1$, the maximum number of ARQ rounds $L$ and the number of fading blocks per ARQ round $B$. As we shall see, for large SNR, the tradeoff expression highlights the role of the asymptotic throughput through $R_1$. Furthermore, the optimal tradeoff expression includes the effect of the space-time spreading dimension of the linear dispersion modulator, providing also a reference of decoding complexity.

We now present the main results of this paper concerning the optimal SNR exponent of ARQ systems.

*Theorem 1:* Consider the channel model (4) with input constellation satisfying the short term average power constraint $\frac{1}{LBT}\mathbb{E}[\|\mathbf{X}\|_F^2] \leq N_t$. The optimal SNR exponent $d^\star(R_1)$ is given by

$$d^\star(R_1) = \begin{cases} N_tN_rLB & \text{for short-term static fading} \\ N_tN_rB & \text{for long-term static fading} \end{cases} \tag{25}$$

Further, this is achieved by Gaussian random codes of rate $R_1 > 0$, provided that the block length is sufficiently long.

*Proof:* Theorem 1 follows immediately as a corollary of [18, Theorem 2] after taking into account the introduction of $B$ in the system. ∎

Theorem 1 states that Gaussian codes achieve maximal diversity gain for any positive rate. As we show in the following, this is not the case with discrete signal constellations $\mathcal{X}_M$. In particular, full diversity is achievable by discrete signal sets provided the rates satisfy $0 \leq R_1 \leq QN_tL$. However, in order to attain full diversity we must restrict the signal constellations to certain properties. In general, due to the discrete nature of these signal sets, a tradeoff between rate, diversity and delay arises. This relationship is expressed in the next theorem.

*Theorem 2:* Consider the channel model (4) satisfying the short term average power constraint $\frac{1}{LBT}\mathbb{E}[\|\mathbf{X}\|_F^2] \leq N_t$, with discrete input signal constellations of cardinality $2^{QN_t}$. The optimal SNR exponent is given by

$$d_D^\star(R_1) = \begin{cases} MN_tN_r\left(1 + \left\lfloor \dfrac{LB}{M}\left(1 - \dfrac{R_1}{LQN_t}\right)\right\rfloor\right) & \text{for short-term static fading} \\[3mm] MN_tN_r\left(1 + \left\lfloor \dfrac{B}{M}\left(1 - \dfrac{R_1}{LQN_t}\right)\right\rfloor\right) & \text{for long-term static fading} \end{cases} \tag{26}$$

over the full range of $0 \leq R_1 \leq QN_tL$ where (26) is continuous.

*Proof (Sketch):* A sketch of the proof is provided here, with the technical details left to Appendix. We first prove the converse and show that the diversity gain $d_D^\star(R_1)$ is upper-bounded by (26). We can use Fano's inequality to show that the outage probability $P_{\text{out}}(\rho, \ell, R_1)$ lower-bounds the error probability $P_e(\rho)$ for a sufficiently large block length. Then we bound the maximum SNR exponent by considering the diversity gain of the outage probability. For large SNR, the instantaneous mutual information is either zero or $QN_t$ bits per channel use, corresponding to when the channel is in deep fade and when the channel is not in deep fade, respectively [26]. Achievability is proved by bounding the error probability of the typical set decoder [18] for ARQ rounds $\ell = 1, \ldots, L-1$, and that of the ML decoder at round $L$, using the union Bhattacharyya bound [40] on a random coded modulation scheme over $\mathcal{Q}$ concatenated with linear dispersion space-time modulation. For finite $T$, we obtain similar conditions to those in [26]. Finally, as $T \to \infty$, we show that the SNR exponent of random codes is given by the Singleton bound for all values of $R_1$ where (26) is continuous. ∎

Theorem 2 states that optimal diversity gain of $N_t N_r L B$ and $N_t N_r B$ for short- and long-term models, respectively, can also be achieved by discrete signal sets coupled with linear dispersion space-time modulators with constellation $\mathcal{X}_B$ ($D = 1$), namely, space-time modulators that spread the symbols of $\mathcal{Q}$ over the $B$ fading blocks at each ARQ round. Under this scenario[3], full diversity is maintained for all rates $0 \leq R_1 \leq QN_t$. However, as anticipated in Section II-B, there is one drawback of practical concern, namely, complexity. In order to achieve full diversity, the linear dispersion space-time modulator needs to spread the symbols of $\mathcal{Q}$ over the $B$ blocks, which implies that the size of the constellation of each ARQ round is $|\mathcal{X}_B| = QN_t BT$. We may, however, choose a modulator that spreads symbols over $M$ blocks where $M < B$ in order to reduce the complexity of the ML decoder. In this case, there is a tradeoff between the parameters of (26). This can be seen as a manifestation of the discrete nature of the input constellation, which limits the performance of the outage probability at high SNR. Theorem 2 generalizes the result of [22] for the quasi-static MIMO channel to the ARQ block-fading case.

The upper bound (26) is also applicable to any systems using block codes over $LB$ independent block-fading periods. The significance of the ARQ framework is that it provides a way of achieving the optimal SNR exponent attained by a block code with $LB$ coded blocks, without always having to transmit all $LB$ code blocks. Indeed, following [18], observe that

$$
\begin{aligned}
\Pr(\mathcal{D}_\ell) &\triangleq \Pr(\mathcal{A}_1^c, \ldots, \mathcal{A}_\ell^c) \\
&\leq \Pr(\mathcal{A}_\ell^c) \\
&= \Pr(\psi_\ell(\widetilde{\mathbf{y}}_\ell, \widetilde{\mathbf{H}}_\ell) = 0) \\
&\leq P_{\text{out}}(\rho, \ell, R_1) + \epsilon \\
&\doteq \rho^{-d_D^*(R_1)}.
\end{aligned}
\tag{27}
$$

---

[3]Within our framework, it would also be possible to modulate over $1 \leq M \leq BL$ periods in the short-term case, namely, spreading the modulation symbols also across ARQ rounds and (26) would remain valid. In particular, letting $M = LB$, we could achieve full diversity over the full range of $R_1$, namely, $0 < R_1 < QN_t L$, which is the same exponent of the Gaussian input. However, generalizing our model to this case, would compress key concepts such as information accumulation in a single formula, rather than the more natural sum expression in (15). In particular, one could define the equivalent channel model up to round $\ell$ as

$$
\widetilde{\mathbf{H}}_\ell = \text{diag}\left(\mathbf{H}_1, \ldots, \mathbf{H}_\ell, \mathbf{0}, \ldots, \mathbf{0}\right) \in \mathbb{C}^{LBN_r \times LBN_t}.
$$

where $\mathbf{0}$ is the zero matrix, and obtain the result.

On substitution of (27) into (13), we find

$$\eta(R_1, L) \overset{\cdot}{\geq} \frac{R_1}{1 + \sum_{\ell=1}^{L-1} \rho^{-d_D^\star(R_1)}} \doteq R_1, \tag{28}$$

which shows that the transmit throughput is asymptotically equal to $R_1$ (since $R_1 \geq \eta(R_1, L)$), the rate of a single ARQ round. In other words, provided the SNR is sufficiently high, ARQ systems which send *on average* $B$ coded blocks can achieve the same diversity gain as that achieved by a block code system which sends $LB$ coded blocks *every time*. This is because in the high SNR regime, most frames can be decoded correctly with high probability based only on the first transmitted code block. ARQ retransmissions are used to correct the rare errors which occur almost exclusively whenever the channel is in outage. While the throughput $\eta(R_1, L)$ is a function of $L$ at mid to low SNR, it converges towards $R_1$ independent of $L$ at sufficiently high SNR. Since the optimal diversity gain is an increasing function of $L$, this behavior can be exploited to increase reliability without suffering code rate losses. However, as noted in [18], this behavior is exhibited only by decoders capable of near perfect error detection (PED). Therefore, the performance of practical error detection schemes can be expected to significantly influence the throughput of ARQ systems.

Since equation (28) relates the asymptotic throughput with the coding parameter $R_1$, the optimal SNR exponent given in (26) gives the *optimal throughput-diversity-delay* tradeoff of MIMO ARQ block-fading channels[4]. Examining the optimal throughput-diversity-delay tradeoff (26) in more detail, we first note that

$$\frac{R_1}{N_t L Q} = \frac{R_0}{Q N_t} = r$$

is the code rate of a binary code. i.e. $0 \leq r \leq 1$, as if the coded modulation scheme $\mathcal{C}_Q$ was obtained itself as the concatenation of a binary code of rate $r$ and length $N_t L Q B T$. Expression (26) implies that the higher we set the target rate $R_1$ (equivalently, $R_0$), the lower the achievable diversity order. In particular, *uncoded* sequences (i.e. $R_1 = Q N_t L$) such as the full diversity modulations [37, 41], achieve optimal diversity gain of $M N_t N_r$, while any code with non-zero $R_1 \leq Q N_t L$ will achieve optimal diversity less than or equal to $M N_t N_r L B$ or $M N_t N_r B$ in the

---

[4]We stress the fact that the coded modulation schemes considered in this paper have a *fixed* rate, and therefore zero multiplexing gain as defined in [17, 18]. However, it is not difficult to show that allowing $Q = \xi \log \rho$ would imply the achievability of the diversity-multiplexing-delay tradeoff of [18].

short- and long-term static models, respectively. This is an intuitively satisfying result as $LB$ and $B$ are precisely the number of independent fading periods in the short- and long-term static models, respectively, each with inherent diversity $MN_tN_r$.

Figures 2, 3, 4 and 5 are graphs of the tradeoff function (26) with varying $Q$, $B$, $L$ and $M$ plotted against the rate of a single ARQ round $R_1$. We show the tradeoff function (26) for both short- and long-term static fading models, respectively.

First we examine the effect of the constellation size $Q$ on the optimal diversity tradeoff function. Figure 2 shows the tradeoff curve for three different values of $Q$. We can see from the plot that the tradeoff curves for higher $Q$ are strictly better than lower $Q$ in terms of achievable diversity gain. This implies that a high order modulation scheme always outperform lower order modulation schemes in the limit of high SNR in terms of error rate performance, for any code rate. Alternatively, a system with high $Q$ can choose to operate at higher code rates than a low $Q$ system and still maintain the same diversity gain.

Figure 3 shows the diversity tradeoff curve for different values of $B$. Similar to the previous tradeoff curve with constellation size $Q$, we observe that systems with high values of $B$ are strictly better than systems with low $B$ (in terms of diversity gain). In addition, we notice that $B$ corresponds to the number of "steps" in the tradeoff function of (26). Systems with low values of $B$ maintain the same diversity gain over wider intervals of rates than systems with high $B$. Relatively, the penalty for using codes with high spectral efficiency is much higher for systems with large $B$ (although these systems will still achieve higher diversity gains than systems with low $B$).

Figure 4 illustrates the effect of the maximum number of allowed ARQ rounds $L$ on the diversity of the system. It is clear from the plot that in the short-term static case the effect of $L$ is to simply shift tradeoff curves upwards. This is intuitively satisfying, since each additional ARQ round represents incremental redundancy, which can be considered as a form of advanced repetition coding. Each additional ARQ round contains $B$ additional independent fading blocks and hence the diversity gain with $L$ ARQ rounds is simply the diversity gain with $L-1$ rounds plus $B$. On the other hand, in the case of long-term static fading, since the different ARQ rounds use the same channel realization, larger $L$ implies a broader range of $R_1$ for which maximum diversity can be achieved.

Figure 5 shows the impact of $M$ on the tradeoff curve. As anticipated in Section II-B, we observe that the larger $M$, the larger the optimal SNR exponent. As $M$ increases, larger diversity is maintained over a larger range of $R_1$. A careful look to (26) reveals that for $M > 1$, each ARQ round behaves as a MIMO block-fading channel with $\frac{B}{M} = D$ blocks, each with inherent diversity $MN_tN_r$, reducing the number of steps of the tradeoff curve. Unfortunately, however, increasing $M$ implies an exponential (in $M$) increase in the overall decoding complexity.

*Remark 1:* In [29, 42], the authors examined the performance of codes over MIMO block-fading channels without ARQ. Using the notation in this paper, the diversity gain based on the worst pairwise error rate performance was shown to be upper-bounded by

$$d_{\text{PEP}}(R) \le N_r \left( 1 + \left\lfloor B \left( N_t - \frac{R}{Q} \right) \right\rfloor \right). \tag{29}$$

The bound in (29) is based on the fact that the rank of a the codeword difference matrix of a given pairwise error event cannot be larger than the minimum number of non-zero rows. The application of the Singleton bound [25] to the minimum number of non-zero rows (interpreted as the Hamming distance of the code) leads the result shown in (29) [29, 42–44]. Since the bound (29) was derived for the non-ARQ case, we will compare it with our results by letting $L = 1$ in (26). An important assumption made in the derivation of (29) is that a signal constellation of cardinality $2^Q$ is used for signaling at each transmit antenna. Under this assumption, the Singleton bound and the rank criterion give rise to the PEP diversity bound (29). In our case, we do not restrict the signals out of each transmit antenna to belong to a constellation of size $2^Q$, but rather, allow for more freedom in the system by linearly modulating (combining) $MN_tT$ $2^Q$-ary symbols to be transmitted over $MT$ channel uses. Figure 6 compares the Singleton bound (29) with our main result (26). As we see, even in the case of $M = 1$ our bound yields a larger exponent. This effect was also observed in [22] for the quasi-static MIMO channel.

## VI. Maximum Distance Separable Space-Time Codes

Having established the main effects of each parameter in (26), we now consider the practical coding aspects of Theorem 2. The diversity tradeoff function (26) can be viewed as a modified version of the Singleton bound [25] with the diversity gain corresponding to the Hamming distance of our code $\mathcal{C}$, viewed as a code of length $\frac{LB}{M} = LD$ constructed over an alphabet of size $2^{QMN_tT}$. This is a useful interpretation and naturally leads us to investigate the role of Singleton-bound-achieving MDS codes. The role of MDS codes as block codes in block-fading channel has been examined extensively in [23, 24, 26, 45].

In this section, we illustrate that the optimal SNR exponent shown in (26) can be achieved with practical MDS coding schemes. The block diagram of the concatenated MIMO ARQ transmitter structure considered in the numerical examples is shown in Figure 7. A codeword of the MDS outer encoder is partitioned into $LB$ blocks. Each such block is then passed through a pseudo-random interleaver, subsequently mapped onto a block of complex symbols according to the signal constellation, and passed through a linear dispersive modulator. In the ARQ transmitter, $B$ blocks of $T$ channel uses are transmitted in each ARQ round. For simplicity, we make use of the MDS convolutional codes presented in [23] to illustrate the practical meaning and importance of the diversity tradeoff curve[5]. The ARQ decoder defined in Section II-C is impractical due to the complexity of the typical set decoder. Instead we develop a bounded-distance ARQ decoder and a sub-optimal iterative *a posteriori* probability (APP) based ARQ decoder, respectively, approximating the behavior of the typical set decoder.

For the numerical examples, we consider two systems. The first system has a maximum number of ARQ rounds of $L = 2$, $B = 1$, and is using the 4-state $[5, 7]_8$ outer convolutional code, while the second system has a maximum number of ARQ rounds of $L = 4$, $B = 1$, and is using the 4-state $[5, 5, 7, 7]_8$ outer convolutional code. The rate of the first ARQ round, $R_1$, is the same for both systems. The two systems are investigated for both single-input, single-output (SISO) and MIMO block-fading channels, subject to short-term static fading and long-term static fading, respectively.

We first consider the use of a bounded-distance ARQ decoder. Define the set of messages $\mathcal{V}_\ell \subseteq \mathcal{M}$, where the corresponding received codeword hypotheses $\widetilde{\mathbf{H}}_\ell \widetilde{\mathbf{X}}_\ell(m)$, $m \in \mathcal{M}$ are within a bounded distance from the received matrix $\widetilde{\mathbf{Y}}_\ell$,

$$\mathcal{V}_\ell \triangleq \left\{ m \in \mathcal{M} \ : \ \left| \widetilde{\mathbf{Y}}_\ell - \widetilde{\mathbf{H}}_\ell \widetilde{\mathbf{X}}_\ell(m) \right|_F^2 \leq \ell B T N_r (1 + \delta) \right\}, \qquad (30)$$

where $\delta > 0$. For $1 \leq \ell \leq L - 1$, the output of the bounded-distance ARQ decoder is then given

---

[5]The main goal of these examples is not to approach the outage probability of the channel, but rather to illustrate the meaning and significance of the results presented in the previous section. If one wants to approach the outage probability, more powerful codes should be employed. For details on outage approaching code ensembles for SISO and MIMO channels the reader is referred to [26, 46–48].

by

$$\psi_\ell(\widetilde{\mathbf{Y}}_\ell, \widetilde{\mathbf{H}}_\ell) = \begin{cases} \tilde{m} & \text{if } \mathcal{V}_\ell = \{\tilde{m}\} \\ 0 & \text{otherwise} \end{cases}. \tag{31}$$

Denoting the true message $\hat{m}$, the undetected error probability is bounded as

$$\Pr(\mathcal{A}_\ell, \mathcal{E}_\ell) = \Pr\left(\bigcup_{\tilde{m} \neq \hat{m}} (\mathcal{V}_\ell = \{\tilde{m}\})\right) \tag{32}$$

$$\leq \Pr\left(|\widetilde{\mathbf{W}}_\ell|_F^2 \geq \ell B T N_r (1+\delta)\right) \tag{33}$$

$$\overset{(a)}{\leq} (1+\delta)^{\ell B T N_r} \exp(-\ell B T N_r \delta), \tag{34}$$

where (a) follows from bounding the chi-squared distribution of $|\widetilde{\mathbf{W}}_\ell|_F^2$ with the Chernoff bound. Finally, letting $\delta = \beta \log \rho$ for $\beta > 0$, we have

$$\Pr(\mathcal{A}_\ell, \mathcal{E}_\ell) \overset{\cdot}{\leq} \rho^{-BTN_r\beta}. \tag{35}$$

This result implies that arbitrarily low undetected error probability can be achieved by the new decoder, at the cost of additional delay. In particular, $\beta$ should be chosen such that $BTN_r\beta \geq d^\star(R_1)$ in order to achieve the optimal ML exponent $d^\star(R_1)$.

Figure 8 illustrates the performance of the two ARQ systems in the short-term SISO static channel. We choose the pseudo-random interleaver to be the trivial identity interleaver, i.e. no interleaving is applied between the outer encoder and the inner modulator. The mapper over $\mathcal{Q}$ is set to be BPSK, the space-time modulation rotation matrix $\mathbf{R} = \mathbf{I}$, and $T = 100$ channel uses. We apply the list Viterbi decoder proposed in [49] to implement the ARQ decoder outlined in (30) and (31). In particular, we choose $\beta = \frac{d^\star(R_1)}{BTN_r}$ to minimize the number of retransmissions.

Considering the $L = 2$ system, the top three curves in Figure 8 show the corresponding outage probability, FER with list decoding and FER with PED. The FER curves are parallel to the outage curve at high SNR, which show that the convolutional MDS codes indeed achieve the optimal diversity gain. The $L = 4$ system corresponds to the bottom three curves of Fig. 8, where again we see that the optimal diversity gain is achieved by the MDS convolutional code.

Comparing the two ARQ systems, it is clear that significant performance gains can be obtained at the expense of higher delays. At FER of $10^{-2}$, the gain of the $L = 4$ system over the $L = 2$ system is already 5 dB. The performance gap increases even more dramatically at higher SNR.

Figure 9 shows the average number of ARQ rounds of the two ARQ systems considered above. For each system, we plot the average number of ARQ rounds with PED, with the list decoder and the lower bound given by (12), respectively. It is clear from the plot that at medium to low SNR, significant loss in throughput is incurred by codes that do not approach the outage probability limit, like convolutional code. Even more loss in throughput is observed when list decoding is used as the error detection mechanism.

Finally, note that the average ARQ round curves converge towards one at high SNR. This agrees with (28) and shows that regardless of the maximum number of allowed ARQ rounds $L$, no spectral efficiency penalties are incurred at sufficiently high SNR. In the limit of high SNR, the transmit throughput $\eta(R_1, L) = R_1$.

Figure 10 and Figure 11 correspond to the error rate and average latency of the same two ARQ systems, under long-term static fading. As predicted by the theoretical results of the previous section, under long-term static fading both schemes have the same SNR exponent. As a matter of fact, despite a 1 dB difference in outage probability, both schemes show virtually the same error probability. As already mentioned in the previous section, in the long-term static case, the ARQ gain translates in a larger range of $R_1$ supported with optimal SNR exponent.

We now consider $2 \times 2$ MIMO systems with $L = 2$ and $L = 4$ using the 4-state $[5, 7]_8$ and $[5, 5, 7, 7]_8$ convolutional codes, concatenated with the optimal $2 \times 2$ linear dispersive modulator suggested in [29]. In this example, the channel coherence time is $T = 32$ channel uses and the mapper over $\mathcal{Q}$ is set to 4QAM. In this case, the bounded-distance ARQ decoder in (31) also becomes impractical, and we therefore resort to sub-optimal iterative error detection and decoding schemes. As a benchmark, we consider an iterative scheme based on the full-complexity APP detector, recursively exchanging code symbol extrinsics with an outer APP decoder, thus generating estimates of the information sequence. Applying the max-log APP detector in place of the full-complexity APP detector provides a low-complexity alternative. For the examples considered here, the full-complexity iterative decoder is roughly twice as complex as the max-log APP alternative. For the full-complexity iterative decoder, we only consider PED as the target benchmark, while for the max-log APP based iterative decoder we consider PED, as well as a non-ideal error detection scheme. At each ARQ round, we run the accumulated received signal through six iterations of the respective iterative detection and decoding algorithms before examining the decoder output.

In the non-ideal error-detection case, errors are detected by examining the soft output of the decoder at each ARQ round. Specifically, we use the minimum bit-reliability criterion [50], checking at the end of each ARQ decoding round whether the minimum bit-wise log-likelihood ratio (LLR) of the information sequence exceeds a threshold, i.e.,

$$\min_{0 < i \leq K}\{|L_{d,i}^{\ell}|\} \geq \theta, \tag{36}$$

where $L_{d,i}^{\ell}$ denotes the $i$th element of the information LLR sequence at the $\ell$th ARQ round and $K$ denotes the length of the LLR vector. If (36) holds, decoding is considered successful, and the information sequence corresponding to the LLR vector is delivered to the sink. The choice of $\theta$ affects both the average latency as well as the error rate of the system. In general, choosing a high $\theta$ encourages the receiver to request additional retransmissions, which in turn reduces the error rate. However, if $\theta$ is set too high, the system behaves as a block coded system and the spectral efficiency advantage of ARQ systems is not realized. Further, it is necessary to increase $\theta$ as a function of SNR in order to achieve error rate performance comparable to that of perfect error detection. To this end, we adjust the threshold as

$$\theta = \max\{1, \beta \log \rho\}, \tag{37}$$

where we have lower bounded $\theta$ in order to encourage retransmissions at low SNR. This choice of $\theta$ was found to perform well when the growth parameter $\beta$ is carefully selected. In the examples shown here, $\beta$ is determined experimentally.

Figure 12 compares the error rate performance of the $L = 2$ system and $L = 4$ system under the short-term fading dynamics. For each system, we plot four curves, corresponding to the lower outage probability bound, obtained by using (24), the PED performance for the two iterative decoders, as well as the minimum bit-reliability criterion (MinLLR) performance for the max-log APP based iterative decoder. In this case we have $\beta = 16$ and $\beta = 32$ for the MinLLR scheme when $L = 2$ and $L = 4$, respectively. We notice that additional retransmissions lead to an appreciable decrease in error rates, and, equally important, the MinLLR criterion performs virtually as good as perfect error detection. Also, we observe no appreciable loss in performance of the max-log APP based iterative decoder as compared to the full-complexity case, confirming the use of the max-log APP approximation is well justified.

Figure 13 compares the average latency (measured in number of ARQ rounds) of the two ARQ systems under the short-term fading scenario. Again, we plot four curves per system, corresponding to the lower bound of expected latency, using (24) and (12), as well as the PED and MinLLR performances. In this case, we observe that the cost of using the MinLLR criterion is mainly an increase in latency, caused by requesting superfluous retransmissions, and again there is no appreciable loss in performance by applying the max-log APP approximation.

Figure 14 and Figure 15 correspond to the error rate and average latency of the same two ARQ systems, under long-term static fading. In this case we have $\beta = 12$ and $\beta = 24$ for the MinLLR scheme when $L = 2$ and $L = 4$, respectively. Once again, as predicted by the theoretical results of the previous section, the error rate curves have the same exponent and, moreover, have very similar gains. Similarly, the advantage of ARQ in this case is that larger throughput can be supported with optimal SNR exponent.

## VII. Conclusions

The focus of this paper is to derive the optimal tradeoff between throughput, diversity gain, and delay for the block-fading MIMO ARQ channel. We prove that for the block-fading MIMO ARQ channel with input constellation satisfying a short-term power constraint, the optimal SNR exponent is given by $N_t N_r L B$ for short-term static fading and $N_t N_r B$ for long-term static fading, which is achieved by Gaussian codes of any positive rate.

When the input signal constellations are constrained to be discrete, this is no longer the case. Due to the discrete nature of these signal sets, a tradeoff between rate, diversity and delay arises. As our main result, we prove that for the block-fading MIMO ARQ channel with discrete input signal constellation of cardinality $2^{QN_t}$ satisfying a short-term power constraint, the optimal SNR exponent is given by a modified Singleton bound, relating all the system parameters. In particular, we show that the tradeoff highlights the roles of the complex-plane signal constellation through $Q$, the rate of the first ARQ round $R_1$, the maximum number of ARQ rounds $L$, and the number of fading blocks per ARQ round $B$. Furthermore, the optimal tradeoff expression includes the effect of the space-time spreading dimension $M$ of the linear dispersion modulator, providing also a reference of decoding complexity.

Finally, we present numerical results demonstrating the practical significance of the theoretical analysis, showing that practical MDS codes achieve the optimal throughput-diversity-delay tradeoff.

## Appendix

In this Appendix, we show the details of the proof of Theorem 2. In particular, we detail the proof for the short-term static model. The proof corresponding to the long-term static model follows exactly the same steps, and it is thus omitted.

### Proof of Theorem 2: Converse

To prove Theorem 2, we first establish the converse and show that the diversity gain is upper-bounded by (26). We assume $N_t \geq N_r$ throughout the analysis with no loss in generality[6].

We start following the arguments in [18, Appendix I] and conclude that by Fano's inequality we can obtain a lower bound to the error probability of the ARQ decoder at any ARQ round $\ell$ by using an ML decoder that operates over the $L$ ARQ rounds. Therefore,

$$P_e(\rho) \geq \mathbb{E}\left[ \left| 1 - \frac{I(\rho|\widetilde{\mathbf{G}}_L)}{R_0 L} - \frac{1}{R_0 L B T} \right|_+ \right] \tag{38}$$

where $|x|_+ = \max\{0, x\}$. Hence, for sufficiently large $T$, we have that [17, 18]

$$P_e(\rho) \dot{\geq} P_{\text{out}}(\rho, L, R_1). \tag{39}$$

Therefore, it follows that we can upper-bound the SNR exponent of the ARQ system by considering the outage probability up to ARQ round $L$.

Now, we study in more detail the properties of $P_{\text{out}}(\rho, L, R_1)$ when discrete signal constellations are used. In particular, we recall that (24) states that

$$I(\rho|\mathbf{G}_\ell) \leq \frac{1}{D} \sum_{d=1}^{D} \min\left\{ QN_t, \frac{1}{M} \sum_{m=1}^{M} \log_2 \det\left( \mathbf{I} + \frac{\rho}{N_t} \mathbf{G}_{\ell,(d-1)M+m} \mathbf{G}_{\ell,(d-1)M+m}^\dagger \right) \right\} \tag{40}$$

and therefore,

$$P_{\text{out}}(\rho, L, R_1)$$

$$\geq \Pr\left( \sum_{\ell=1}^{L} \frac{1}{D} \sum_{d=1}^{D} \min\left\{ QN_t, \frac{1}{M} \sum_{m=1}^{M} \log_2 \det\left( \mathbf{I} + \frac{\rho}{N_t} \mathbf{G}_{\ell,(d-1)M+m} \mathbf{G}_{\ell,(d-1)M+m}^\dagger \right) \right\} < R_1 \right) \tag{41}$$

$$= \Pr\left( \sum_{\ell=1}^{L} \sum_{d=1}^{D} \min\left\{ QN_t, \frac{1}{M} \sum_{m=1}^{M} \sum_{i=1}^{N_r} \log_2 \left( 1 + \frac{\rho}{N_t} \lambda_{\ell,(d-1)M+m,i} \right) \right\} < DR_1 \right) \tag{42}$$

---

[6]If $N_t < N_r$, it suffices to replace $\det\left( \mathbf{I} + \mathbf{G}_{\ell,b}\mathbf{G}_{\ell,b}^\dagger \right)$ by $\det\left( \mathbf{I} + \mathbf{G}_{\ell,b}^\dagger \mathbf{G}_{\ell,b} \right)$ in the computation of the input-output mutual information with Gaussian inputs and all the arguments still follow.

where $\lambda_{\ell,(d-1)M+m,1} \leq \ldots \leq \lambda_{\ell,(d-1)M+m,N_r}$ are the ordered $N_r$ eigenvalues of the $N_r \times N_r$ matrix $\mathbf{G}_{\ell,(d-1)M+m}\mathbf{G}_{\ell,(d-1)M+m}^{\dagger}$ corresponding to ARQ round $\ell$ and fading block $(d-1)M+m$.

We now characterize the behavior of the outage probability at high SNR. Following [17] we define the *SNR normalized* eigenvalues as

$$\alpha_{\ell,b,i} \triangleq -\frac{\log \lambda_{\ell,b,i}}{\log \rho}. \tag{43}$$

The joint probability distribution of $\boldsymbol{\alpha}_{\ell,b} = (\alpha_{\ell,b,1}, \ldots, \alpha_{\ell,b,M})$, can be described using a result in [17, Lemma 3]

$$f(\boldsymbol{\alpha}_{\ell,b}) = K_{N_t,N_r}^{-1}(\log \rho)^{N_r} \prod_{i=1}^{N_r} \rho^{-(N_t-N_r+1)\alpha_{\ell,b,i}} \prod_{i<j} \left(\rho^{-\alpha_{\ell,b,i}} - \rho^{-\alpha_{\ell,b,j}}\right)^2 \exp\left(-\sum_{i=1}^{N_r} \rho^{-\alpha_{\ell,b,i}}\right), \tag{44}$$

where $K_{N_t,N_r}$ is a normalizing constant. Then it follows that

$$P_{\text{out}}(\rho, L, R_1)$$

$$\geq \Pr\left(\sum_{\ell=1}^{L}\sum_{d=1}^{D} \min\left\{QN_t, \frac{1}{M}\sum_{m=1}^{M}\sum_{i=1}^{N_r} \log_2\left(1 + \frac{\rho}{N_t}\lambda_{\ell,(d-1)M+m,i}\right)\right\} < DR_1\right) \tag{45}$$

$$\doteq \Pr\left(\sum_{\ell=1}^{L}\sum_{d=1}^{D} \min\left\{QN_t, \frac{1}{M}\sum_{m=1}^{M}\sum_{i=1}^{N_r} \log_2 \rho^{\left|1-\alpha_{\ell,(d-1)M+m,i}\right|_+}\right\} < DR_1\right) \tag{46}$$

$$\doteq \Pr\left(\sum_{\ell=1}^{L}\sum_{d=1}^{D} \min\left\{QN_t, \frac{\log_2 \rho}{M}\sum_{m=1}^{M}\sum_{i=1}^{N_r} \left|1-\alpha_{\ell,(d-1)M+m,i}\right|_+\right\} < DR_1\right). \tag{47}$$

If we now define

$$\widetilde{\boldsymbol{\alpha}}_{\ell,d} \triangleq \left(\boldsymbol{\alpha}_{\ell,(d-1)M+1}', \ldots, \boldsymbol{\alpha}_{\ell,dM}'\right)' \in \mathbb{R}^{MN_r} \tag{48}$$

$$= \left(\alpha_{\ell,(d-1)M+1,1}, \ldots, \alpha_{\ell,(d-1)M+1,N_r}, \ldots, \alpha_{\ell,dM,1}, \ldots, \alpha_{\ell,dM,N_r}\right)' \tag{49}$$

equation (47) becomes

$$P_{\text{out}}(\rho, L, R_1) \dot{\geq} \Pr\left(\sum_{\ell=1}^{L}\sum_{d=1}^{D} (1 - \mathbb{1}\{\widetilde{\boldsymbol{\alpha}}_{\ell,d} \succeq \mathbf{1}\}) < \frac{DR_1}{QN_t}\right) \tag{50}$$

where $\mathbf{a} \succeq \mathbf{b}$ denotes componentwise inequality, i.e., $a_i \geq b_i, \forall i = 1, \ldots, n$ for some $\mathbf{a}, \mathbf{b} \in \mathbb{R}^n$ and $\mathbf{1}$ is the all-one vector, since

$$\min\left\{QN_t, \frac{\log_2 \rho}{M}\sum_{m=1}^{M}\sum_{i=1}^{N_r} \left|1-\alpha_{\ell,(d-1)M+m,i}\right|_+\right\} \doteq \begin{cases} 0 & \text{when } \boldsymbol{\alpha}_{\ell,d} \succeq \mathbf{1} \\ QN_t & \text{otherwise.} \end{cases} \tag{51}$$

                                                                                    

This means that asymptotically for large SNR, when all the components of $\widetilde{\boldsymbol{\alpha}}_{k,d}$ are larger or equal than one (deep fades) the mutual information tends to $0$, and to $QN_t$ otherwise. Following similar steps as in [17] we can write that

$$P_{\text{out}}(\rho, L, R_1) \dot{\geq} \int_{\boldsymbol{\alpha} \in \widetilde{\mathcal{O}}_L \cap \mathbb{R}_+^{LDMN_r}} \exp\left(-\log \rho \sum_{\ell=1}^{L} \sum_{d=1}^{D} \sum_{m=1}^{M} \sum_{i=1}^{N_r} (2i-1+N_t-N_r)\alpha_{\ell,(d-1)M+m,i}\right) d\boldsymbol{\alpha} \tag{52}$$

where the large SNR outage event is given by

$$\widetilde{\mathcal{O}}_L = \left\{ \boldsymbol{\alpha} \in \mathbb{R}^{LDMN_r} \ : \ \sum_{\ell=1}^{L} \sum_{d=1}^{D} \left(1 - \mathbb{1}\{\widetilde{\boldsymbol{\alpha}}_{\ell,d} \succeq \mathbf{1}\}\right) < \frac{DR_1}{QN_t} \right\} \tag{53}$$

$$= \left\{ \boldsymbol{\alpha} \in \mathbb{R}^{LDMN_r} \ : \ \sum_{\ell=1}^{L} \sum_{d=1}^{D} \mathbb{1}\{\widetilde{\boldsymbol{\alpha}}_{\ell,d} \succeq \mathbf{1}\} > D\left(L - \frac{R_1}{QN_t}\right) \right\} \tag{54}$$

and $\boldsymbol{\alpha} \triangleq \left(\widetilde{\boldsymbol{\alpha}}'_{1,1}, \ldots, \widetilde{\boldsymbol{\alpha}}'_{L,D}\right)' \in \mathbb{R}^{LDMN_r}$. Applying Varadhan's lemma [51] we have that

$$d_D^\star(R_1) \leq \inf_{\boldsymbol{\alpha} \in \widetilde{\mathcal{O}}_L \cap \mathbb{R}_+^{LDMN_r}} \left\{ \sum_{\ell=1}^{L} \sum_{d=1}^{D} \sum_{m=1}^{M} \sum_{i=1}^{N_r} (2i-1+N_t-N_r)\alpha_{\ell,(d-1)M+m,i} \right\}. \tag{55}$$

The infimum (55) is solved by considering two cases. If $R_1 > LQN_t$, then the infimum is satisfied by $\widetilde{\boldsymbol{\alpha}}_{\ell,d} = \mathbf{0}$ for all $\ell$ and $d$, hence the diversity gain is zero. Alternatively, if $R_1 \leq LQN_t$, then among all possible vectors $\widetilde{\boldsymbol{\alpha}}_{\ell,d}$, for $\ell = 1, \ldots, L$ and $d = 1, \ldots, D$, we need to have $k$ vectors equal to the all-ones vector ($\widetilde{\boldsymbol{\alpha}}_{\ell,d} = \mathbf{1}$), for some $k \in \mathbb{Z}$ in order to satisfy the infimum. The condition to be met is written

$$k > D\left(L - \frac{R_1}{QN_t}\right), \tag{56}$$

which implies that in order to achieve the infimum $k$ should be

$$k = 1 + \left\lfloor D\left(L - \frac{R_1}{QN_t}\right) \right\rfloor. \tag{57}$$

Since $\sum_{m=1}^{M} \sum_{i=1}^{N_r} 2i - 1 + N_t - N_r = MN_tN_r$, we upper-bound the optimal SNR exponent as

$$d_D^\star(R_1) \leq MN_tN_r \left(1 + \left\lfloor \frac{LB}{M}\left(1 - \frac{R_1}{LQN_t}\right) \right\rfloor\right), \tag{58}$$

which proves the desired converse result.

## Proof of Theorem 2: Achievability

To prove the achievability of the upper-bound on the SNR exponent in (58), we examine the average frame error rate obtained using random codes and the ARQ decoder described in Section II-C. This decoder behaves like a typical set decoder for ARQ rounds $\ell = 1, \ldots, L-1$, and as an ML decoder at round $L$ [38]. Since the channel matrix $\widetilde{\mathbf{H}}_L$ encompasses the channel realizations of all ARQ rounds, with a slight abuse of notation we can express the error probability conditioned on the fading realization as

$$P_e(\rho|\widetilde{\mathbf{H}}_L) = \sum_{\ell=1}^{L-1} \Pr\left( \mathcal{D}_{\ell-1}, \bigcup_{\substack{\hat{m} \neq m \\ \hat{m} \neq 0}} \psi_\ell(\widetilde{\mathbf{Y}}_\ell, \widetilde{\mathbf{H}}_\ell) = \hat{m} \right) + \Pr\left( \mathcal{D}_{L-1}, \bigcup_{\hat{m} \neq m} \psi_L(\widetilde{\mathbf{Y}}_L, \widetilde{\mathbf{H}}_L) = \hat{m} \right)$$

(59)

where all parameters are defined in Section III. As shown in [18, 38, Appendix I], $\forall \delta > 0$ and sufficiently large $T$, there exists a code for which the error probability corresponding to the first $L-1$ rounds can be bounded as

$$\sum_{\ell=1}^{L-1} \Pr\left( \mathcal{D}_{\ell-1}, \bigcup_{\substack{\hat{m} \neq m \\ \hat{m} \neq 0}} \psi_\ell(\widetilde{\mathbf{Y}}_\ell, \widetilde{\mathbf{H}}_\ell) = \hat{m} \right) < \delta.$$

(60)

Therefore,

$$P_e(\rho|\mathbf{H}_L) \leq (L-1)\delta + P_e^{\mathrm{ml}}(\rho|\mathbf{H}_L)$$

(61)

where

$$P_e^{\mathrm{ml}}(\rho|\mathbf{H}_L) \triangleq \Pr\left( \mathcal{D}_{L-1}, \bigcup_{\hat{m} \neq m} \psi_L(\widetilde{\mathbf{Y}}_L, \widetilde{\mathbf{H}}_L) = \hat{m} \right)$$

(62)

is the error probability of an ML decoding error at the $L$th ARQ round. We now characterize the behavior of $P_e^{\mathrm{ml}}(\rho|\mathbf{H}_L)$ for a particular code construction[7], namely, random codes constructed

---

[7]We could simply conclude the proof by following the same arguments of the proof in [18, Appendix I], namely, using

$$P_e^{\mathrm{ml}}(\rho|\mathbf{H}_L) < \delta + \mathbb{1}\left\{\mathbf{H}_L \in \mathcal{O}_L\right\}$$

to argue that $P_e(\rho) \dot{\leq} \rho^{d_D^\star(R_1)}$ (see [18, Appendix I] for details). However, the specific analysis of the ML decoding error probability for round $L$ using random codes encompasses the standard quasistatic and block-fading MIMO channels with no ARQ as special cases, and therefore is of broader interest.

over $\mathcal{Q}$, concatenated with random linear dispersion space-time modulators described in Section II-B.

Following the steps of [22] we consider that the $2^{R_0 LBT}$ codewords of $\mathcal{C}_{\mathcal{Q}}$ are generated with the uniform probability distribution over $\mathcal{Q}$, namely, $\forall \mathbf{c}_{\mathcal{Q}} \in \mathcal{C}_{\mathcal{Q}}$,

$$p(\mathbf{c}_{\mathcal{Q}}) = \prod_{k=1}^{LBTN_t} \frac{1}{|\mathcal{Q}|} = \frac{1}{2^{QLBTN_t}}. \tag{63}$$

Each codeword $\mathbf{c}_{\mathcal{Q}} \in \mathcal{C}_{\mathcal{Q}}$ is partitioned into $LD$ vectors, denoted $\mathbf{c}_{\mathcal{Q},\ell,d} \in \mathcal{Q}^{MTN_t}$, where $\ell = 1, \ldots, L$ and $d = 1, \ldots, D$, such that $\mathbf{c}_{\mathcal{Q}} = [\mathbf{c}_{\mathcal{Q},1,1}, \ldots, \mathbf{c}_{\mathcal{Q},L,D}]'$. Now let

$$\mathcal{R} = \left\{ \mathbf{R} \in \mathbb{R}^{MTN_t \times MTN_t} : \mathbf{R}\mathbf{R}' = \mathbf{R}'\mathbf{R} = \mathbf{I} \right\} \tag{64}$$

denote the set of orthogonal matrices of dimension $MTN_t \times MTN_t$. As outlined in Section II-B, the modulated signals are given by

$$\widehat{\mathbf{x}}_{\ell,d} = \mathbf{R}\mathbf{c}_{\mathcal{Q},\ell,d}. \tag{65}$$

Then, if we define

$$\widehat{\mathbf{X}}_{\ell,d} \overset{\Delta}{=} \mathrm{mat}_{MN_t \times T}(\widehat{\mathbf{x}}_{\ell,d}) \tag{66}$$

where the operator $\mathbf{A} = \mathrm{mat}_{n \times m}(\mathbf{a})$ formats vector $\mathbf{a} \in \mathbb{C}^{nm}$ into an $n \times m$ matrix, we have that the portion of codeword transmitted over ARQ round $\ell$ can be written as

$$\mathbf{X}_{\ell} = \left[ \widehat{\mathbf{X}}'_{\ell,1}, \ldots, \widehat{\mathbf{X}}'_{\ell,D} \right]'. \tag{67}$$

Then we have that the conditional pairwise error probability is given by

$$P\left( \mathbf{X}(n) \to \mathbf{X}(k) \mid \widetilde{\mathbf{H}}_L = \widetilde{\mathbf{G}}_L \right) = Q\left( \sqrt{\frac{\rho}{2N_t} \left\| \widetilde{\mathbf{G}}_L(\mathbf{X}(n) - \mathbf{X}(k)) \right\|_F^2} \right) \tag{68}$$

$$\leq \exp\left( -\frac{\rho}{4N_t} \left\| \widetilde{\mathbf{G}}_L(\mathbf{X}(n) - \mathbf{X}(k)) \right\|_F^2 \right). \tag{69}$$

It follows from the structure of $\widetilde{\mathbf{G}}_L$ that

$$P\left( \mathbf{X}(n) \to \mathbf{X}(k) \mid \widetilde{\mathbf{H}}_L = \widetilde{\mathbf{G}}_L \right) \leq \prod_{\ell=1}^{L} \prod_{d=1}^{D} \exp\left( -\frac{\rho}{4N_t} \left\| \widehat{\mathbf{G}}_{\ell,d}(\widehat{\mathbf{x}}_{\ell,d}(n) - \widehat{\mathbf{x}}_{\ell,d}(k)) \right\|^2 \right) \tag{70}$$

$$= \prod_{\ell=1}^{L} \prod_{d=1}^{D} \exp\left( -\frac{\rho}{4N_t} \sum_{m=1}^{M} \left\| \mathbf{G}_{\ell,(d-1)M+m}(\mathbf{X}_{\ell,(d-1)M+m}(n) - \mathbf{X}_{\ell,(d-1)M+m}(k)) \right\|_F^2 \right). \tag{71}$$

If the elements of $\mathcal{R}$ are drawn with the uniform probability distribution, it follows from [52, Theorem 1] that $\mathbf{R}$ has full diversity with probability one, namely, the matrices $\mathbf{X}_{\ell,(d-1)M+m}(n) -$ $\mathbf{X}_{\ell,(d-1)M+m}(k)$ have full rank[8]. We now apply the singular value decomposition (SVD) [36] to both channel and difference matrices

$$\mathbf{G}_{\ell,(d-1)M+m} = \mathbf{U}\boldsymbol{\Lambda}^{\frac{1}{2}}_{\ell,(d-1)M+m}\mathbf{V}^{\dagger} \tag{72}$$

and

$$\mathbf{X}_{\ell,(d-1)M+m}(n) - \mathbf{X}_{\ell,(d-1)M+m}(k) = \mathbf{A}\mathbf{D}^{\frac{1}{2}}_{\ell,(d-1)M+m}\mathbf{B}^{\dagger} \tag{73}$$

and get that

$$P\left(\mathbf{X}(n) \rightarrow \mathbf{X}(k) \mid \widetilde{\mathbf{H}}_L = \widetilde{\mathbf{G}}_L\right)$$

$$\leq \prod_{\ell=1}^{L}\prod_{d=1}^{D}\exp\left(-\frac{\rho}{4N_t}\sum_{m=1}^{M}\left\|\mathbf{G}_{\ell,(d-1)M+m}(\mathbf{X}_{\ell,(d-1)M+m}(n) - \mathbf{X}_{\ell,(d-1)M+m}(k))\right\|_F^2\right) \tag{74}$$

$$= \prod_{\ell=1}^{L}\prod_{d=1}^{D}\exp\left(-\frac{\rho}{4N_t}\sum_{m=1}^{M}\left\|\mathbf{U}\boldsymbol{\Lambda}^{\frac{1}{2}}_{\ell,(d-1)M+m}\mathbf{V}^{\dagger}\mathbf{A}\mathbf{D}^{\frac{1}{2}}_{\ell,(d-1)M+m}\mathbf{B}^{\dagger}\right\|_F^2\right) \tag{75}$$

$$= \prod_{\ell=1}^{L}\prod_{d=1}^{D}\exp\left(-\frac{\rho}{4N_t}\sum_{m=1}^{M}\left\|\boldsymbol{\Lambda}^{\frac{1}{2}}_{\ell,(d-1)M+m}\mathbf{P}\mathbf{D}^{\frac{1}{2}}_{\ell,(d-1)M+m}\right\|_F^2\right), \tag{76}$$

where $\mathbf{P} = \mathbf{V}^{\dagger}\mathbf{A}$ is unitary. The diagonal matrices $\boldsymbol{\Lambda}^{\frac{1}{2}}_{\ell,(d-1)M+m}$ and $\mathbf{D}^{\frac{1}{2}}_{\ell,(d-1)M+m}$ are composed of the singular values of the channel matrix $\mathbf{G}_{\ell,(d-1)M+m}$ and codeword difference matrix $\mathbf{X}_{\ell,(d-1)M+m}(n) - \mathbf{X}_{\ell,(d-1)M+m}(k)$, respectively. As mentioned earlier, the matrices $\mathbf{X}_{\ell,(d-1)M+m}(n) - \mathbf{X}_{\ell,(d-1)M+m}(k)$ have full rank with probability one, which implies that the the $MN_r$ singular values in $\mathbf{D}^{\frac{1}{2}}_{\ell,(d-1)M+m}$ are all non-zero for $m = 1, \ldots, M$, $d = 1, \ldots, D$ and $\ell = 1, \ldots, L$. If we now define

$$\boldsymbol{\Gamma}_{\ell,(d-1)M+m} \overset{\Delta}{=} \mathbf{P}\mathbf{D}_{\ell,(d-1)M+m}\mathbf{P}^{\dagger} \tag{77}$$

and

$$\boldsymbol{\gamma}_{\ell,(d-1)M+m} \overset{\Delta}{=} \mathrm{diag}\left(\boldsymbol{\Gamma}_{\ell,(d-1)M+m}\right) \tag{78}$$

$$= (\gamma_{\ell,(d-1)M+m,1}, \ldots, \gamma_{\ell,(d-1)M+m,N_r}), \tag{79}$$

---

[8]As it will be clear in the following, random rotations are not essential in the proof. It is sufficient to rely on the existence of a particular $\mathbf{R}$ with full diversity [29, 32–35, 37, 41].

we can rewrite (76) as

$$P\left(\mathbf{X}(n) \rightarrow \mathbf{X}(k) \mid \widetilde{\mathbf{H}}_L = \widetilde{\mathbf{G}}_L\right) \leq \prod_{\ell=1}^{L} \prod_{d=1}^{D} \exp\left(-\sum_{m=1}^{M} \sum_{i=1}^{N_r} \frac{\gamma_{\ell,(d-1)M+m,i}}{4N_t} \rho^{1-\alpha_{\ell,(d-1)M+m,i}}\right). \tag{80}$$

Averaging (80) over the code ensemble, namely $\mathbf{X}(n), \mathbf{X}(k)$ and $\mathbf{R}$, we get that

$$\overline{P\left(\mathbf{X}(n) \rightarrow \mathbf{X}(k) | \boldsymbol{\alpha}\right)} = \prod_{\ell=1}^{L} \prod_{d=1}^{D} \frac{1}{2^{QMTN_t}}\left[1\right.$$

$$\left. + \frac{1}{2^{QMTN_t}} \sum_{\mathbf{c}_{\mathcal{Q},\ell,d}(n) \neq \mathbf{c}_{\mathcal{Q},\ell,d}(k)} \mathbb{E}_{\mathbf{R}}\left[\exp\left(-\sum_{m=1}^{M} \sum_{i=1}^{N_r} \frac{\gamma_{\ell,(d-1)M+m,i}}{4N_t} \rho^{1-\alpha_{\ell,(d-1)M+m,i}}\right)\right]\right]. \tag{81}$$

If we now sum over the $2^{R_0 BLT}$ codewords, we have the union bound

$$\overline{P_e(\rho | \boldsymbol{\alpha})} \leq 2^{R_0 BLT} \prod_{\ell=1}^{L} \prod_{d=1}^{D} \frac{1}{2^{QMTN_t}}\left[1\right.$$

$$\left. + \frac{1}{2^{QMTN_t}} \sum_{\mathbf{c}_{\mathcal{Q},\ell,d}(n) \neq \mathbf{c}_{\mathcal{Q},\ell,d}(k)} \mathbb{E}_{\mathbf{R}}\left[\exp\left(-\sum_{m=1}^{M} \sum_{i=1}^{N_r} \frac{\gamma_{\ell,(d-1)M+m,i}}{4N_t} \rho^{1-\alpha_{\ell,(d-1)M+m,i}}\right)\right]\right] \tag{82}$$

$$= \exp\left(-LDMTQN_t \log(2)\left[1 - \frac{R_0}{QN_t} - \frac{1}{LDMTQN_t} \sum_{\ell=1}^{L} \sum_{d=1}^{D} \log_2\left(1\right.\right.\right. \tag{83}$$

$$\left.\left.\left. + \frac{1}{2^{QMTN_t}} \sum_{\mathbf{c}_{\mathcal{Q},\ell,d}(n) \neq \mathbf{c}_{\mathcal{Q},\ell,d}(k)} \mathbb{E}_{\mathbf{R}}\left[\exp\left(-\sum_{m=1}^{M} \sum_{i=1}^{N_r} \frac{\gamma_{\ell,(d-1)M+m,i}}{4N_t} \rho^{1-\alpha_{\ell,(d-1)M+m,i}}\right)\right]\right)\right]\right) \tag{84}$$

$$= \exp\left(-LDMTQN_t \log(2) E(\rho, \boldsymbol{\alpha})\right) \tag{85}$$

where we have defined the union bound exponent as

$$E(\rho, \boldsymbol{\alpha}) \triangleq 1 - \frac{R_0}{QN_t} - \frac{1}{LDMTQN_t} \sum_{\ell=1}^{L} \sum_{d=1}^{D} \log_2\left(1\right. \tag{86}$$

$$\left. + \frac{1}{2^{QMTN_t}} \sum_{\mathbf{c}_{\mathcal{Q},\ell,d}(n) \neq \mathbf{c}_{\mathcal{Q},\ell,d}(k)} \mathbb{E}_{\mathbf{R}}\left[\exp\left(-\sum_{m=1}^{M} \sum_{i=1}^{N_r} \frac{\gamma_{\ell,(d-1)M+m,i}}{4N_t} \rho^{1-\alpha_{\ell,(d-1)M+m,i}}\right)\right]\right). \tag{87}$$

Following similar arguments to those in [22, 26], we use the dominated convergence theorem

[53] to obtain that

$$\lim_{\rho \to \infty} \mathbb{E}_{\mathbf{R}} \left[ \exp \left( -\sum_{m=1}^{M} \sum_{i=1}^{N_r} \frac{\gamma_{\ell,(d-1)M+m,i}}{4N_t} \rho^{1-\alpha_{\ell,(d-1)M+m,i}} \right) \right]$$

$$= \mathbb{E}_{\mathbf{R}} \left[ \lim_{\rho \to \infty} \exp \left( -\sum_{m=1}^{M} \sum_{i=1}^{N_r} \frac{\gamma_{\ell,(d-1)M+m,i}}{4N_t} \rho^{1-\alpha_{\ell,(d-1)M+m,i}} \right) \right] \tag{88}$$

$$= 1 - \mathbb{1}\{\widetilde{\boldsymbol{\alpha}}_{\ell,d} \succ \mathbf{1}\}, \tag{89}$$

since $\gamma_{\ell,(d-1)M+m,i} > 0$ with probability one. For $\epsilon > 0$ and large SNR, the union bound exponent $E(\rho, \boldsymbol{\alpha})$ can be lower-bounded by

$$E_\epsilon(\rho, \boldsymbol{\alpha}) \triangleq 1 - \frac{R_0}{QN_t} - \frac{1}{LD} \sum_{\ell=1}^{L} \sum_{d=1}^{D} \mathbb{1}\{\widetilde{\boldsymbol{\alpha}}_{\ell,d} \succeq \mathbf{1} - \boldsymbol{\epsilon}\}. \tag{90}$$

Let now

$$\mathcal{E}_\epsilon = \left\{ \boldsymbol{\alpha} \in \mathbb{R}^{LDMN_r} : E_\epsilon(\rho, \boldsymbol{\alpha}) \leq 0 \right\} \tag{91}$$

$$= \left\{ \boldsymbol{\alpha} \in \mathbb{R}^{LDMN_r} : \sum_{\ell=1}^{L} \sum_{d=1}^{D} \mathbb{1}\{\widetilde{\boldsymbol{\alpha}}_{\ell,d} \succeq \mathbf{1} - \boldsymbol{\epsilon}\} \geq LD \left( 1 - \frac{R_0}{QN_t} \right) \right\}. \tag{92}$$

Then we can bound the overall error probability as

$$\overline{P_e(\rho)} \dot{\leq} \int_{\boldsymbol{\alpha} \in \mathbb{R}^{LDMN_r}} \min \left\{ 1, \exp \left( -LDMTQN_t \log(2) E_\epsilon(\rho, \boldsymbol{\alpha}) \right) \right\} p(\boldsymbol{\alpha}) d\boldsymbol{\alpha}. \tag{93}$$

In a similar way to what it is done in [26], we consider codes with block length $T(\rho)$ such that

$$\tau \triangleq \lim_{\rho \to \infty} \frac{T(\rho)}{\log \rho}. \tag{94}$$

That is, we consider sufficiently long codewords large SNR such that the error probability is never dominated by the event when two codewords coincide. Thus, we can write that,

$$\overline{P_e(\rho)} \dot{\leq} \int_{\boldsymbol{\alpha} \in \mathcal{E}_\epsilon \cap \mathbb{R}_+^{LDMN_r}} \exp \left( -\log \rho \sum_{\ell=1}^{L} \sum_{d=1}^{D} \sum_{m=1}^{M} \sum_{i=1}^{N_r} (2i - 1 + N_t - N_r) \alpha_{\ell,(d-1)M+m,i} \right) d\boldsymbol{\alpha}$$

$$+ \int_{\boldsymbol{\alpha} \in \mathcal{E}_\epsilon^c \cap \mathbb{R}_+^{LDMN_r}} \exp \left( -\log \rho \sum_{\ell=1}^{L} \sum_{d=1}^{D} \sum_{m=1}^{M} \sum_{i=1}^{N_r} (2i - 1 + N_t - N_r) \alpha_{\ell,(d-1)M+m,i} \right.$$

$$\left. + \tau LDMQN_t \log(2) E_\epsilon(\rho, \boldsymbol{\alpha}) \right) d\boldsymbol{\alpha} \tag{95}$$

and therefore, the random coding exponent is lower-bounded by

$$d^{(r)}(R_1) \geq \sup_{\epsilon > 0} \min\{d_1, d_2\} \tag{96}$$

where

$$d_1 = \inf_{\boldsymbol{\alpha} \in \mathcal{E}_\epsilon \cap \mathbb{R}_+^{LDMN_r}} \left\{ \sum_{\ell=1}^{L} \sum_{d=1}^{D} \sum_{m=1}^{M} \sum_{i=1}^{N_r} (2i - 1 + N_t - N_r)\alpha_{\ell,(d-1)M+m,i} \right\} \tag{97}$$

is the exponent for large enough codewords and

$$d_2 = \inf_{\boldsymbol{\alpha} \in \mathcal{E}_\epsilon^c \cap \mathbb{R}_+^{LDMN_r}} \left\{ \sum_{\ell=1}^{L} \sum_{d=1}^{D} \sum_{m=1}^{M} \sum_{i=1}^{N_r} (2i - 1 + N_t - N_r)\alpha_{\ell,(d-1)M+m,i} \right.$$

$$\left. + \tau LDMQN_t \log(2) E_\epsilon(\rho, \boldsymbol{\alpha}) \right\} \tag{98}$$

$$= \inf_{\boldsymbol{\alpha} \in \mathcal{E}_\epsilon^c \cap \mathbb{R}_+^{LDMN_r}} \left\{ \tau LDMQN_t \log(2) \left( 1 - \frac{R_0}{QN_t} \right) \right.$$

$$\left. + M(N_tN_r - \tau QN_t \log(2)) \sum_{\ell=1}^{L} \sum_{d=1}^{D} \mathbb{1}\{\boldsymbol{\alpha}_{\ell,d} \succeq \mathbf{1} - \boldsymbol{\epsilon}\} \right\} \tag{99}$$

is the exponent that characterizes the finite block length.

Following similar steps to those in the converse, the SNR exponent of the first component $d_1$ can be written

$$d_1 \geq (1 - \epsilon)MN_tN_r \left\lceil \frac{LB}{M} \left( 1 - \frac{R_1}{LQN_t} \right) \right\rceil . \tag{100}$$

Following similar arguments as in [26], we see that if $0 \leq \tau QN_t \log(2) < N_tN_r$ then the infimum (99) is given by

$$LDM\tau QN_t \log(2) \left( 1 - \frac{R_0}{QN_t} \right) . \tag{101}$$

Otherwise, if $\tau QN_t \log(2) \geq N_tN_r$, then the infimum is

$$\tau LDMQN_t \log(2) \left( 1 - \frac{R_0}{QN_t} \right) + M(N_tN_r(1 - \epsilon) - \tau QN_t \log(2)) \left( \left\lceil LD \left( 1 - \frac{R_0}{QN_t} \right) \right\rceil - 1 \right) . \tag{102}$$

The random coding SNR exponent lower-bound can be tightened by letting $\epsilon \to 0$. By collecting the results together, we see that for sufficiently large $\tau$, $d_2$ coincides with $d_1$. In fact, one observes that for $T \to \infty$, the overall error probability is given by the probability of the event $\mathcal{E}_\epsilon$, since the second integral in (95) vanishes. Hence the diversity lower-bound coincides with the diversity upper-bound (26) for all rates except at the discontinuities.

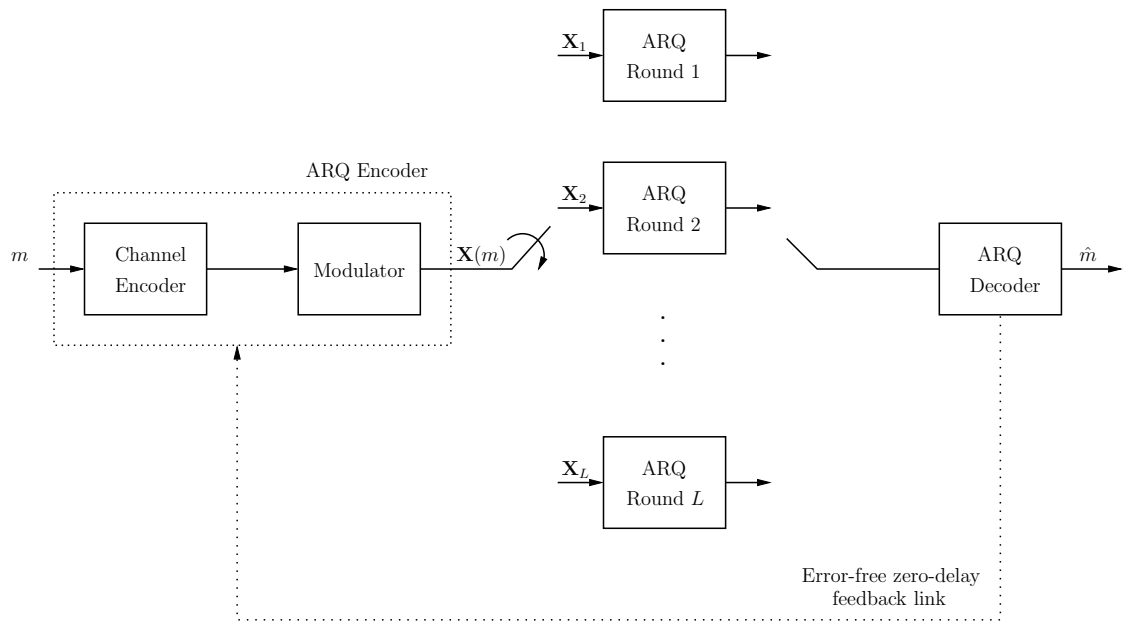

(a) Block diagram of ARQ.

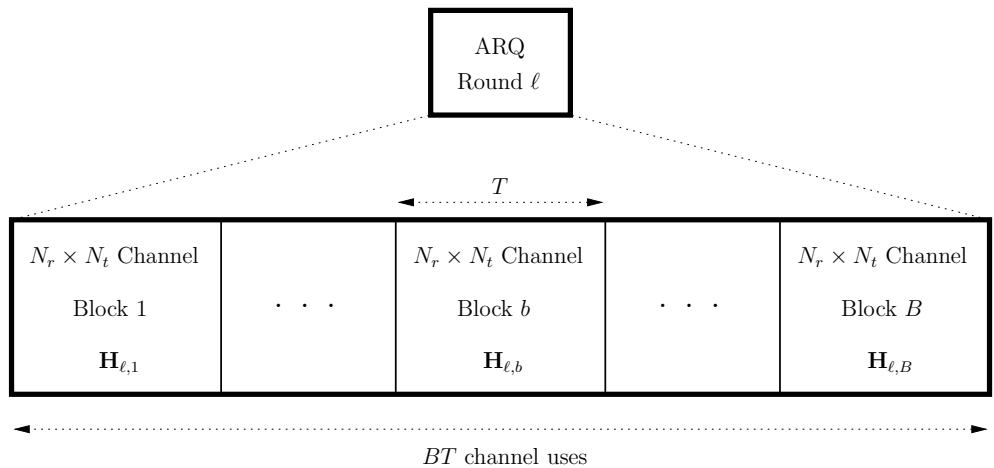

(b) Channel at ARQ round $\ell$.

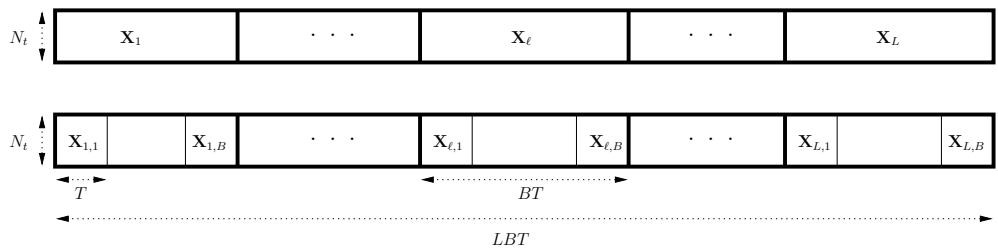

(c) Codeword structure.

Fig. 1.  MIMO ARQ system model.

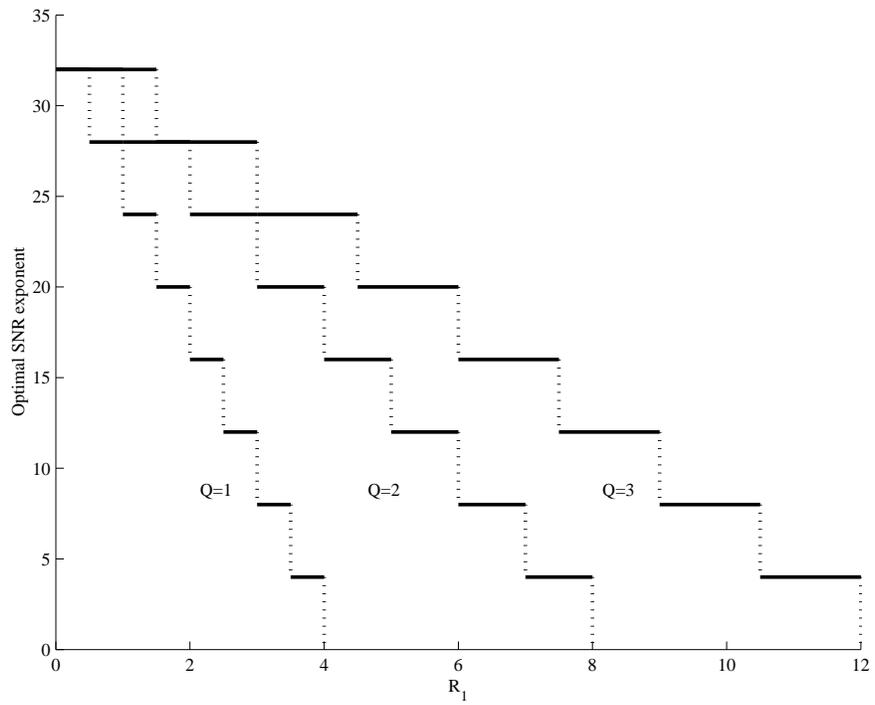

(a) Short-term static fading.

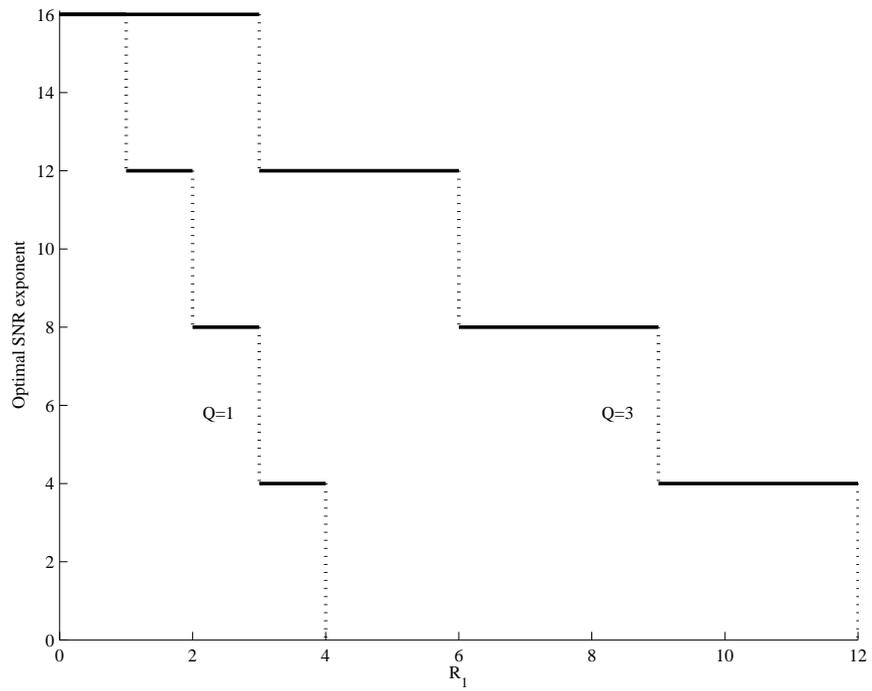

(b) Long-term static fading.

Fig. 2. Optimal diversity tradeoff curve corresponding to $L = 2, B = 4, M = 1$ for a $2 \times 2$ MIMO channel.

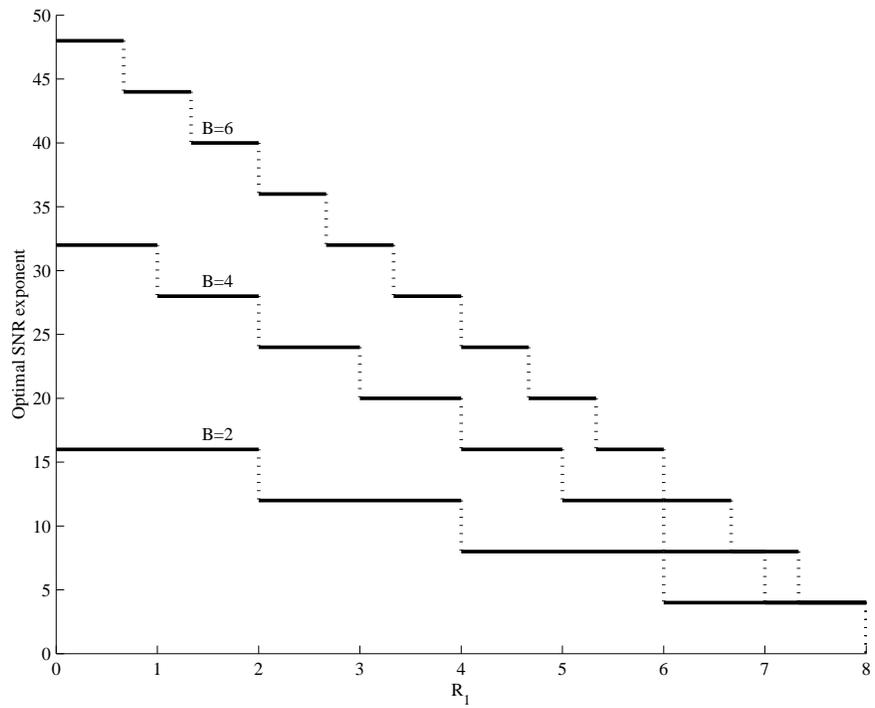

(a) Short-term static fading.

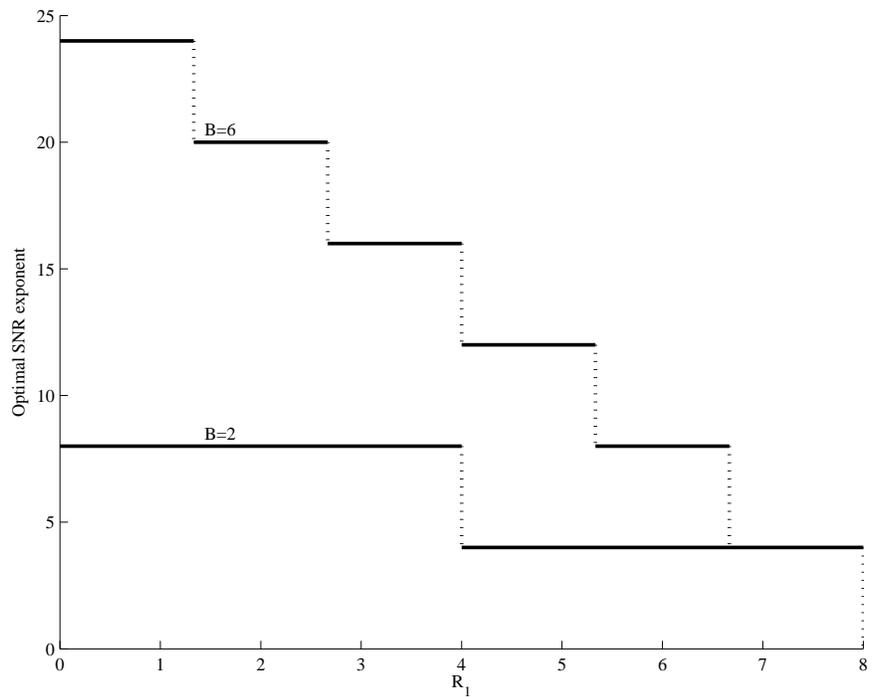

(b) Long-term static fading.

Fig. 3.   Optimal diversity tradeoff curve corresponding to $L = 2, Q = 2, M = 1$ for a $2 \times 2$ MIMO channel.

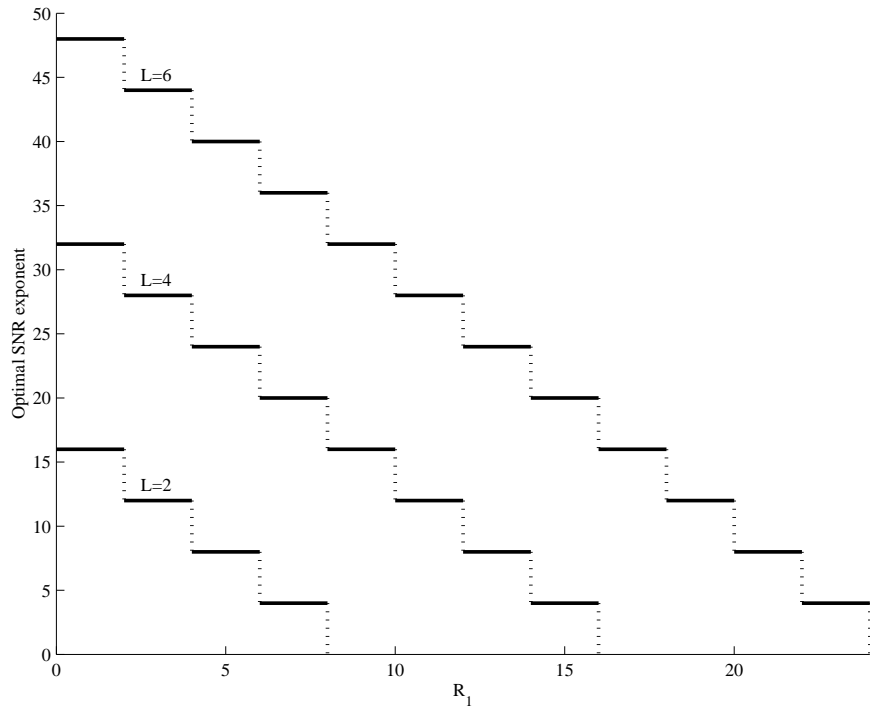

(a) Short-term static fading.

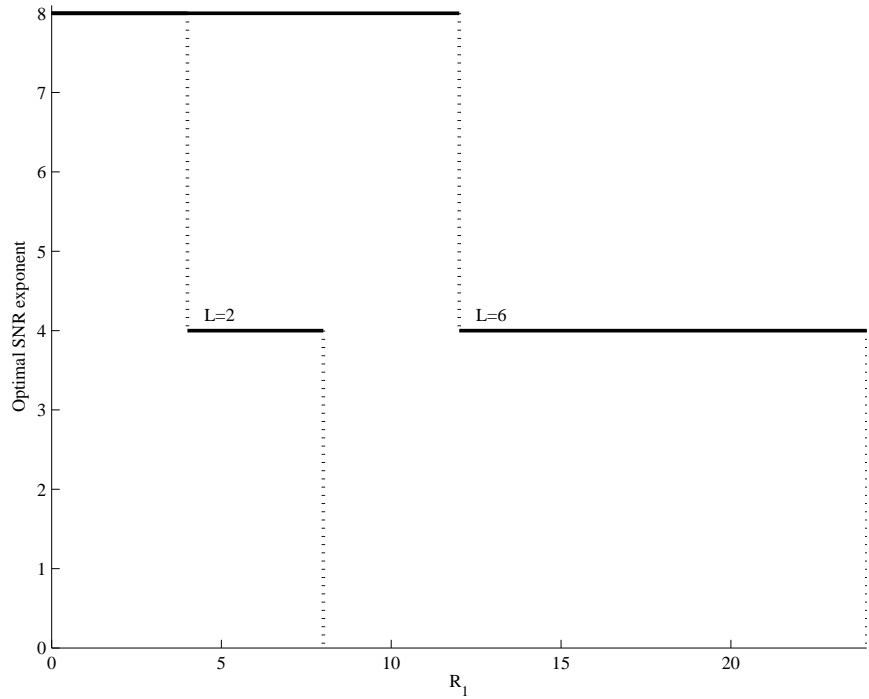

(b) Long-term static fading.

Fig. 4. Optimal diversity tradeoff curve corresponding to $B = 2, Q = 2, M = 1$ for a $2 \times 2$ MIMO channel.

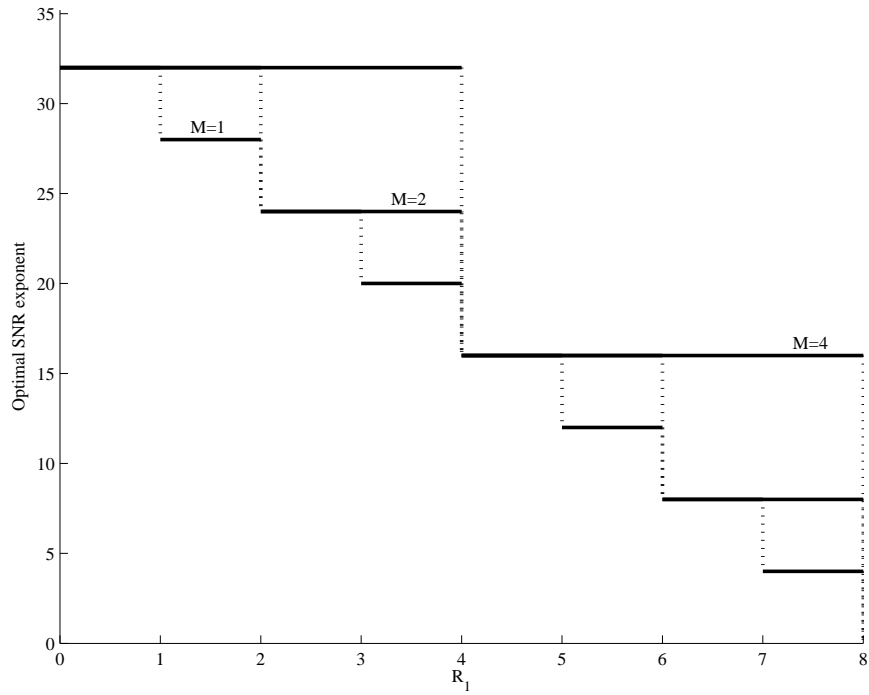

(a) Short-term static fading.

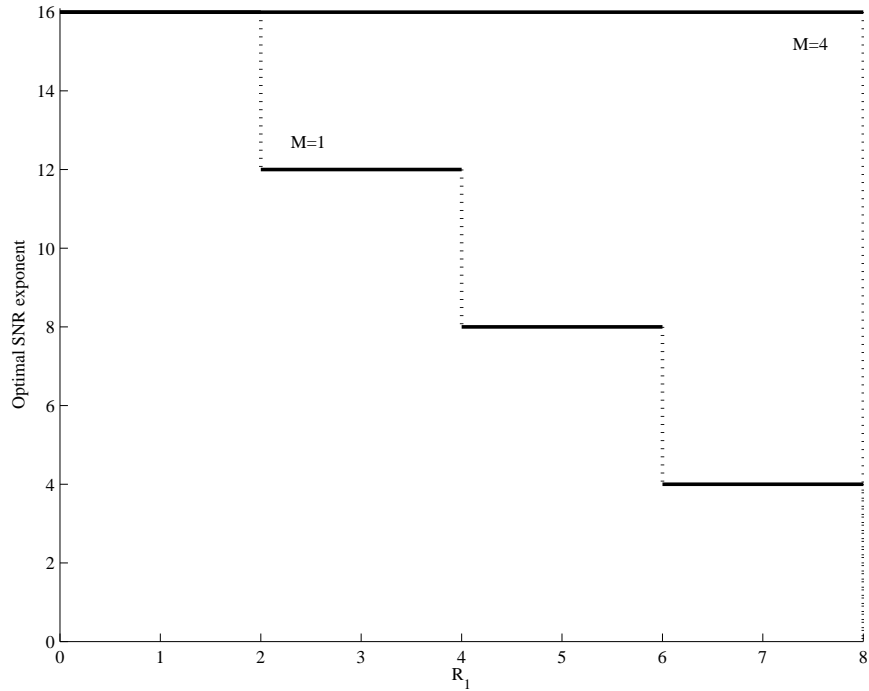

(b) Long-term static fading.

Fig. 5. Optimal diversity tradeoff curve corresponding to $B = 4, Q = 2, L = 2$ for a $2 \times 2$ MIMO channel. The curves with 8 steps correspond to $M = 1$, those with 4 to $M = 2$ and those with 2 to $M = B = 2$.

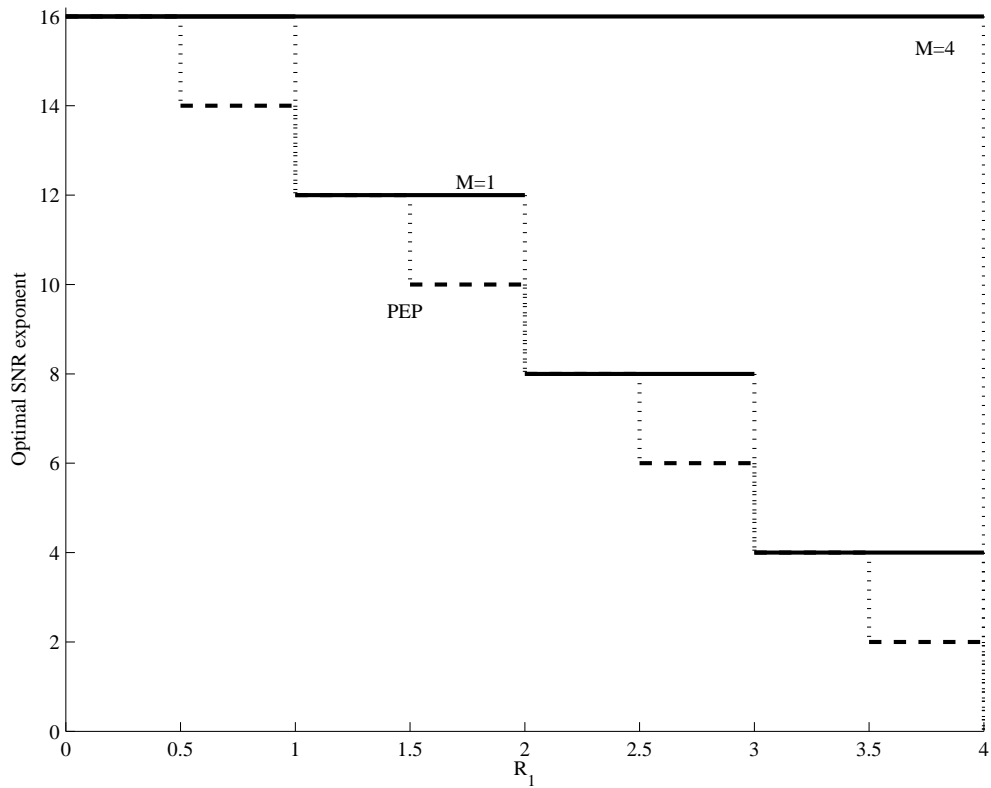

Fig. 6. Comparison of the optimal diversity tradeoff curve for with (29) (dashed steps) for $B = 4, Q = 2, L = 1$ and $M = 1, 4$ in a $2 \times 2$ MIMO channel.

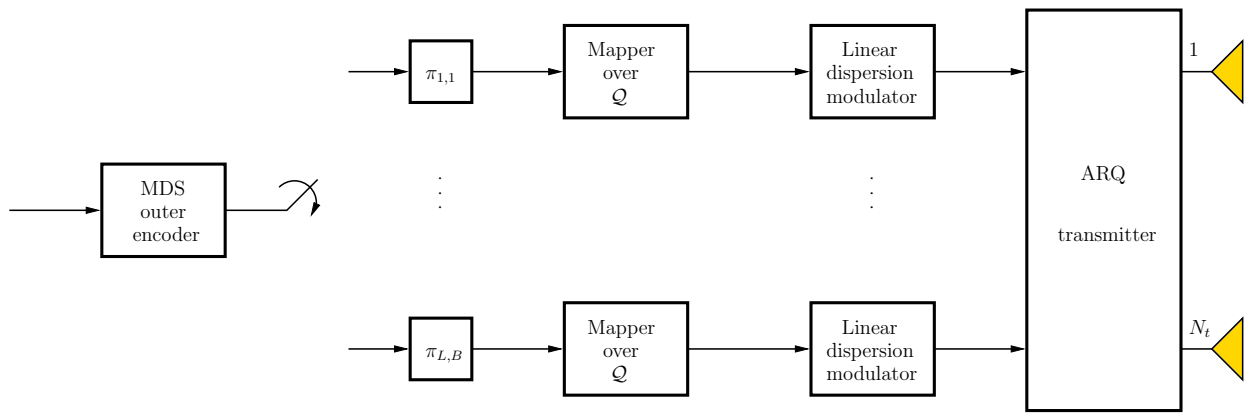

Fig. 7. Block diagram of the concatenated MIMO ARQ architecture. The interleaver corresponding to ARQ round $\ell$ and fading block $b$ is denoted by $\pi_{\ell,b}$.

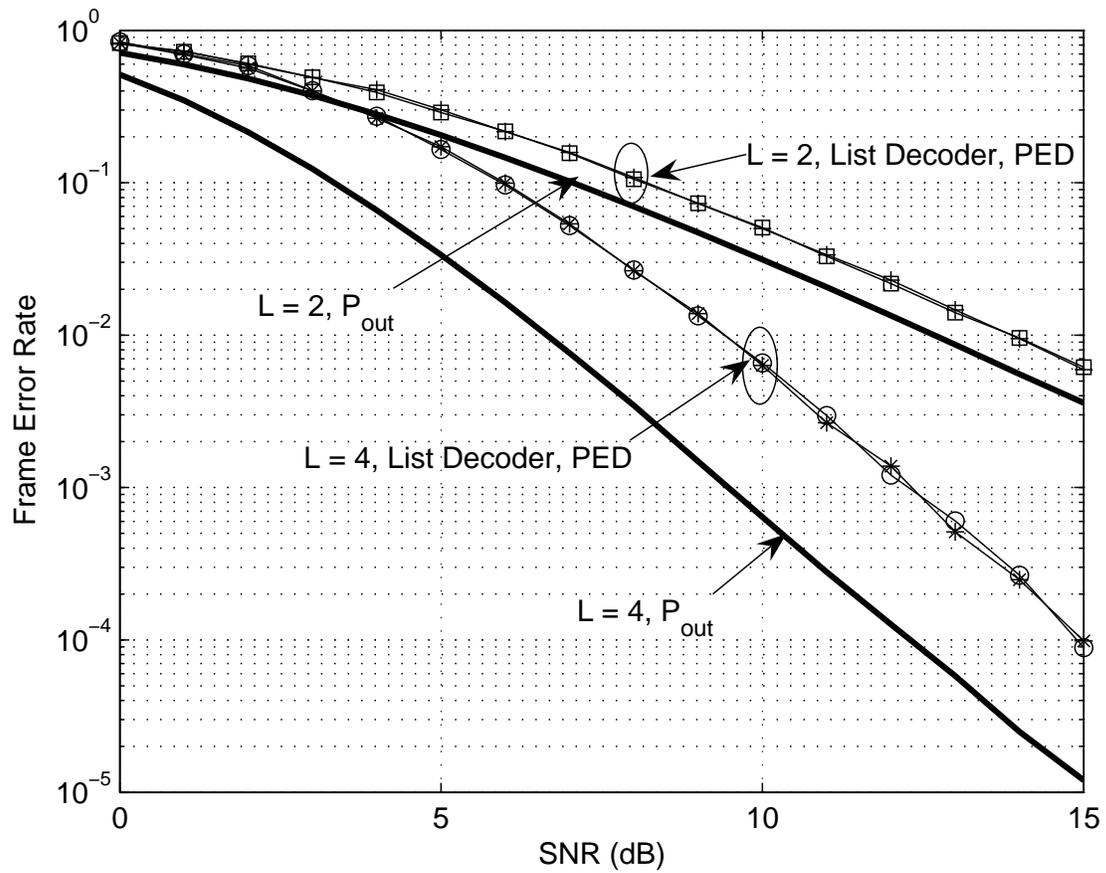

Fig. 8.  FER with MDS convolutional code over a short-term static SISO channel corresponding to $B = 1$, $Q = 1$ and $T = 100$.

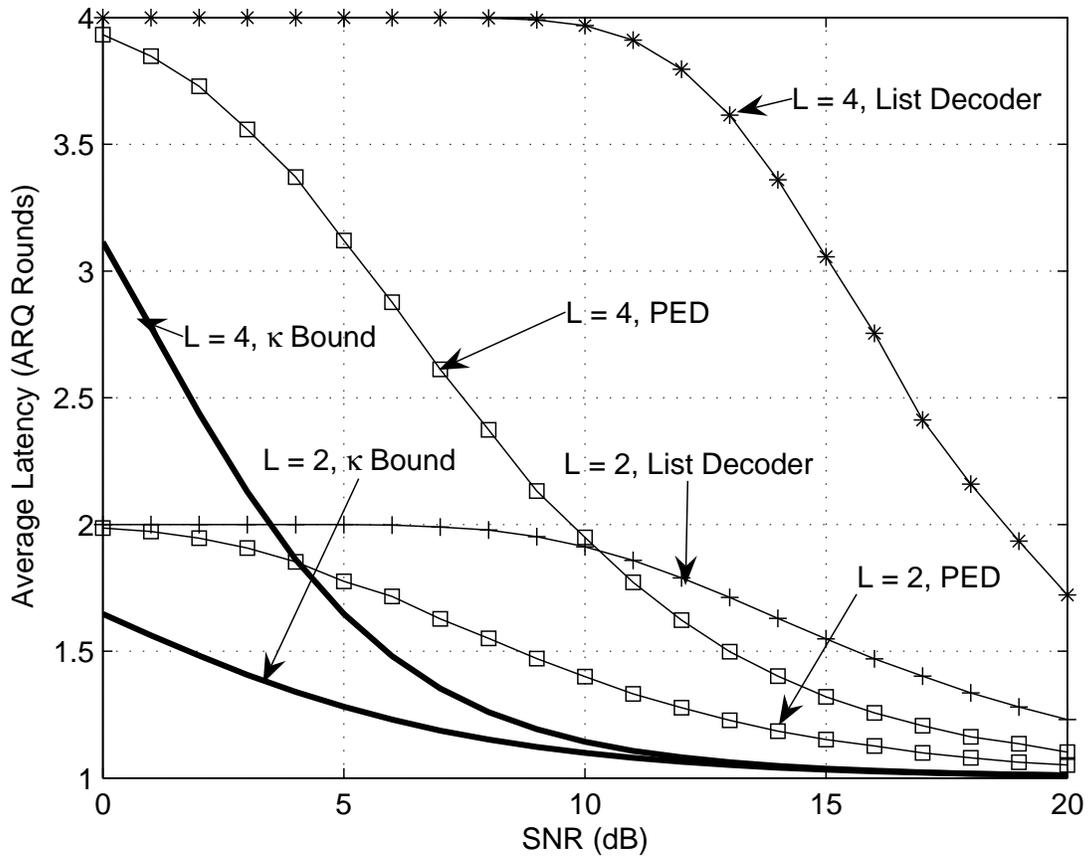

Fig. 9. Average number of ARQ rounds for MDS convolutional codes over a short-term static SISO channel corresponding to $B = 1$, $Q = 1$ and $T = 100$.

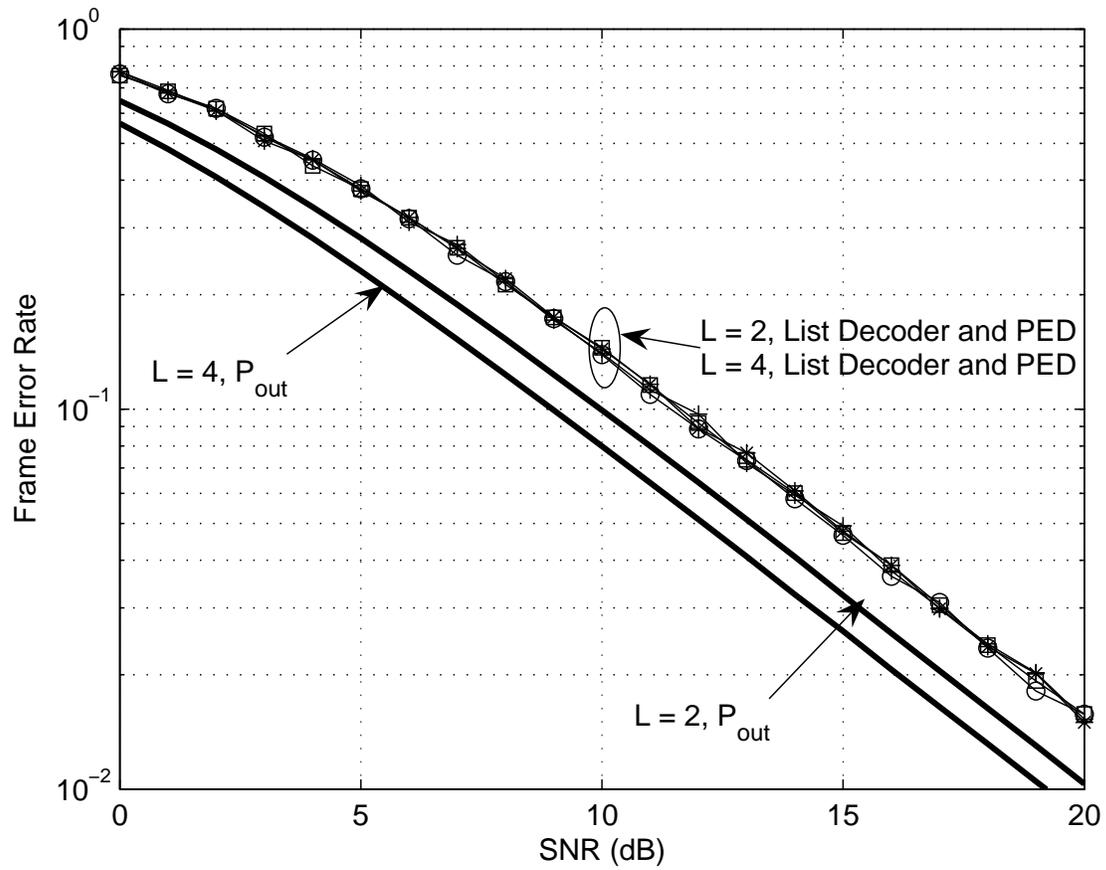

Fig. 10. FER with MDS convolutional code over a long-term static SISO channel corresponding to $B = 1$, $Q = 1$ and $T = 100$.

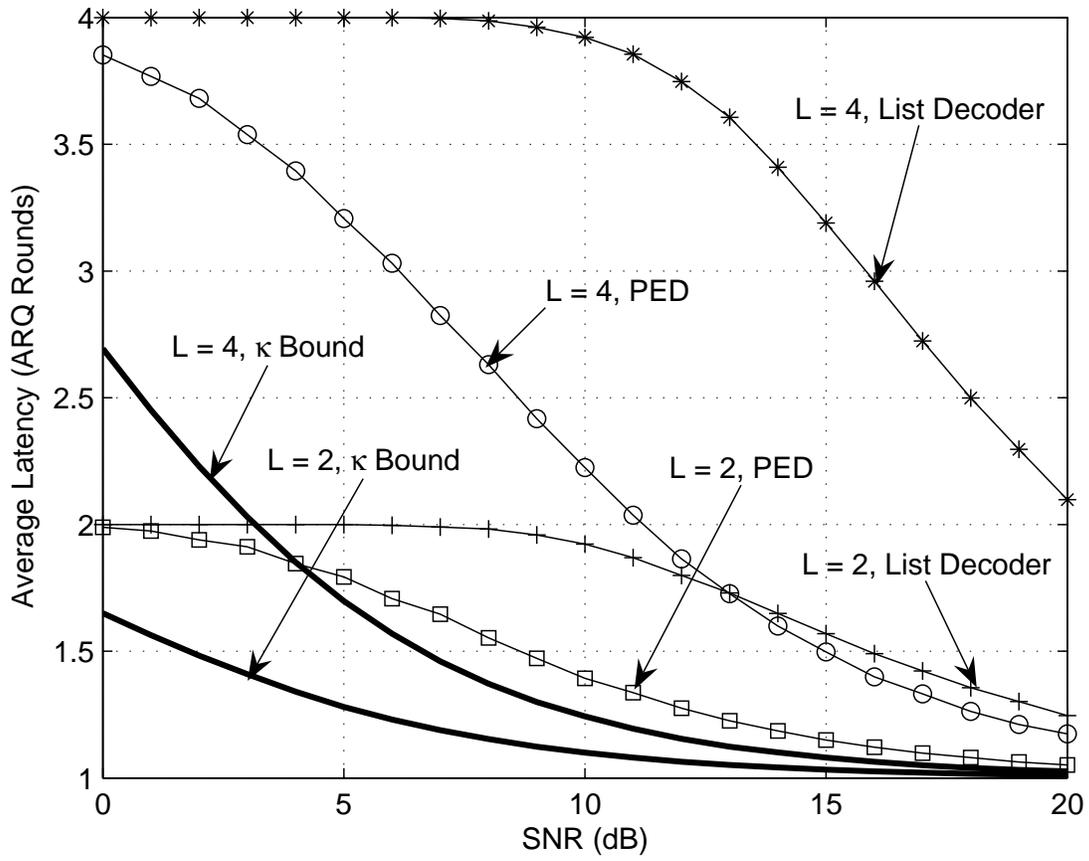

Fig. 11. Average number of ARQ rounds for MDS convolutional codes over a long-term static SISO channel corresponding to $B = 1$, $Q = 1$ and $T = 100$.

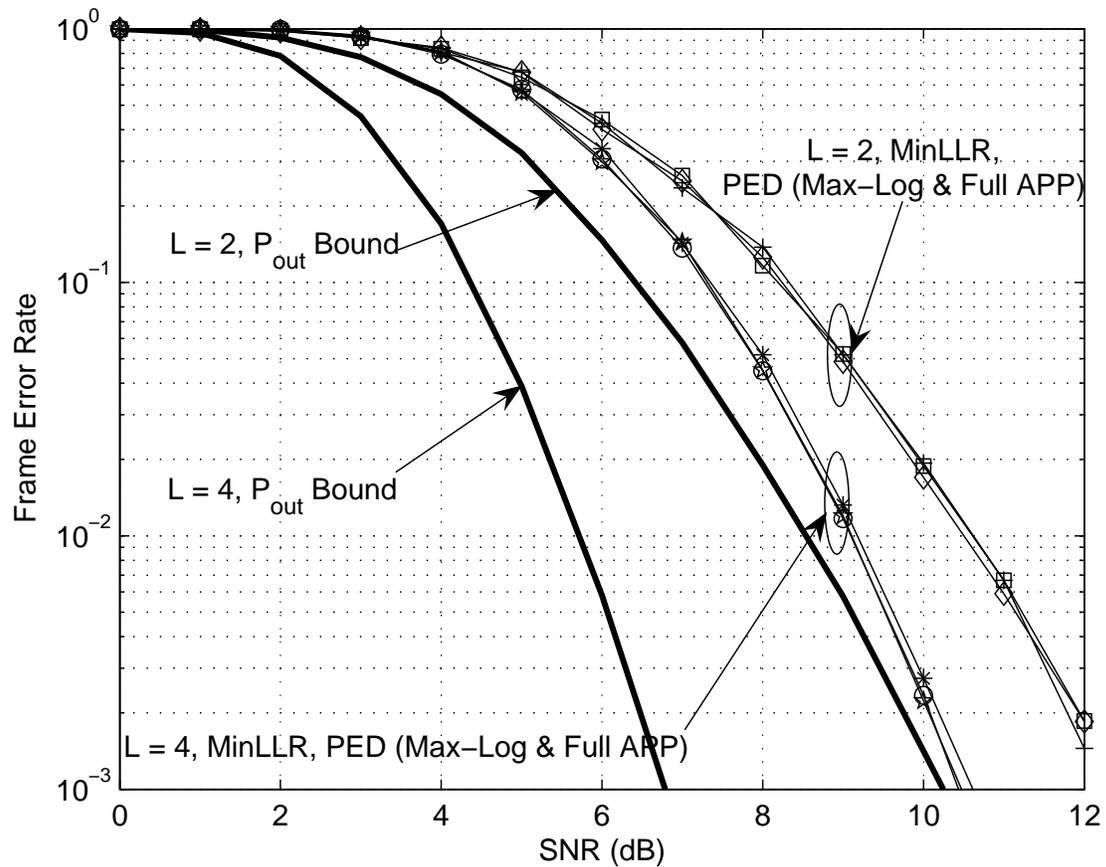

Fig. 12. FER with MDS convolutional code over a short-term static $2 \times 2$ MIMO channel corresponding to $B = 1$, $Q = 2$ and $T = 32$. The thick solid lines are the lower outage probability bounds. For $L = 2$, diamonds correspond to full-complexity APP detection with PED, while squares and crosses correspond to max-log APP detection with PED and MinLLR, respectively. For $L = 4$, pentagrams correspond to full-complexity APP detection with PED, while circles and asterisks correspond to max-log APP detection with PED and MinLLR, respectively.

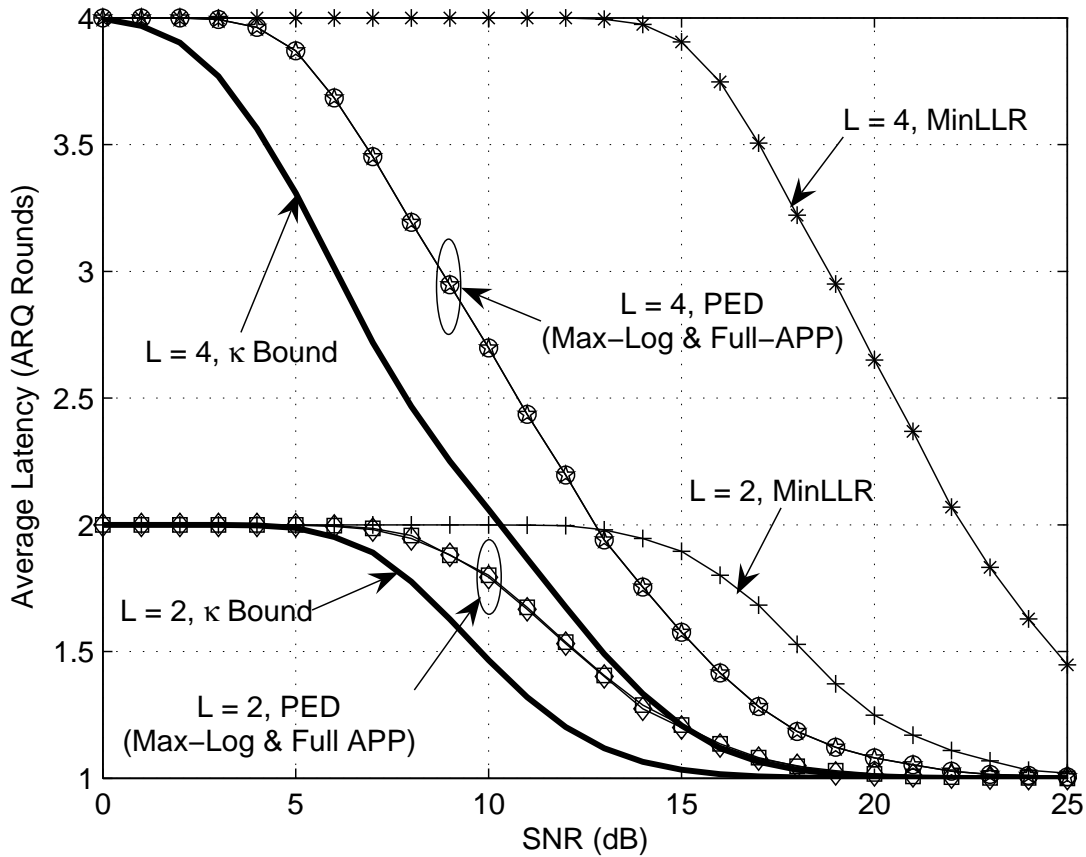

Fig. 13. Average number of ARQ rounds for MDS convolutional codes over a short-term static $2 \times 2$ MIMO channel corresponding to $B = 1$, $Q = 2$ and $T = 32$. The thick solid lines are the lower bounds on expected latency. For $L = 2$, diamonds correspond to full-complexity APP detection with PED, while squares and crosses correspond to max-log APP detection with PED and MinLLR, respectively. For $L = 4$, pentagrams correspond to full-complexity APP detection with PED, while circles and asterisks correspond to max-log APP detection with PED and MinLLR, respectively.

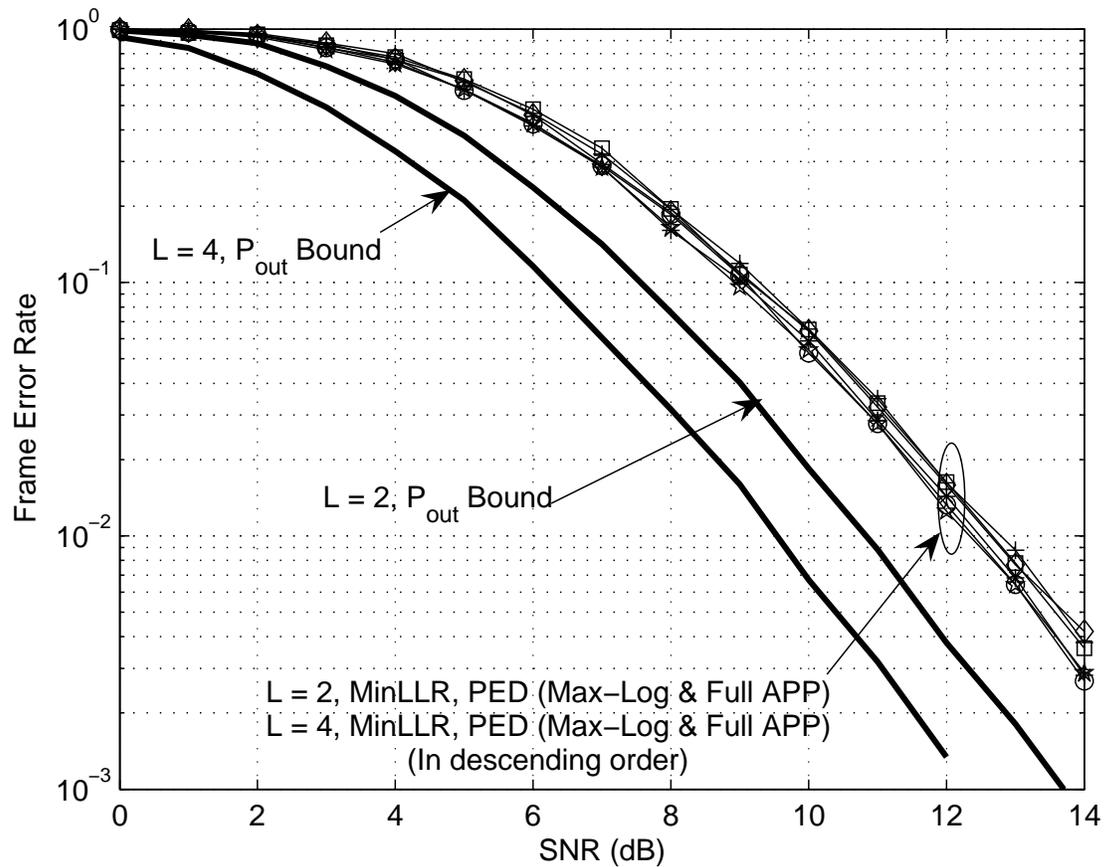

Fig. 14. FER with MDS convolutional code over a long-term static $2 \times 2$ MIMO channel corresponding to $B = 1$, $Q = 2$ and $T = 32$. The thick solid lines are the lower outage probability bounds. For $L = 2$, diamonds correspond to full-complexity APP detection with PED, while squares and crosses correspond to max-log APP detection with PED and MinLLR, respectively. For $L = 4$, pentagrams correspond to full-complexity APP detection with PED, while circles and asterisks correspond to max-log APP detection with PED and MinLLR, respectively.

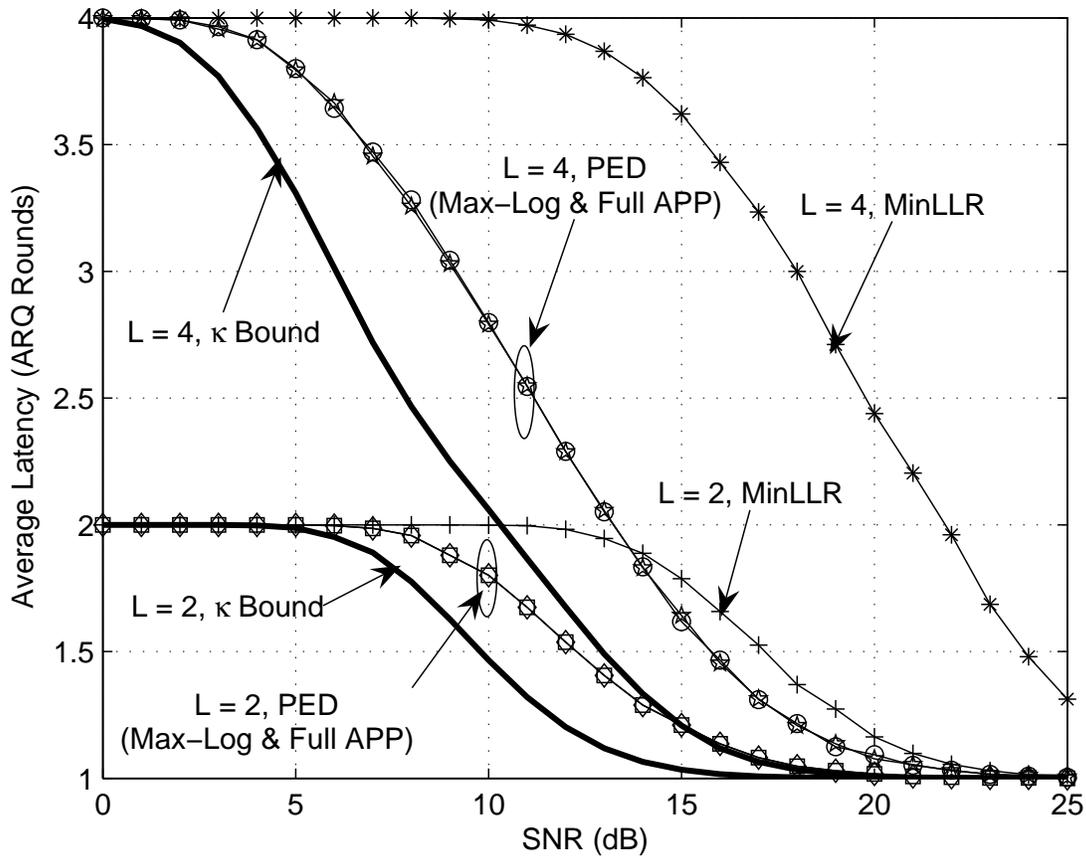

Fig. 15. Average number of ARQ rounds for MDS convolutional codes over a long-term static $2 \times 2$ MIMO channel corresponding to $B = 1$, $Q = 2$ and $T = 32$. The thick solid lines are the lower bounds on expected latency. For $L = 2$, diamonds correspond to full-complexity APP detection with PED, while squares and crosses correspond to max-log APP detection with PED and MinLLR, respectively. For $L = 4$, pentagrams correspond to full-complexity APP detection with PED, while circles and asterisks correspond to max-log APP detection with PED and MinLLR, respectively.